\documentclass[aps,pra,twocolumn,showpacs,superscriptaddress,longbibliography]{revtex4-1}
\usepackage{graphicx} %Include figure files,superscriptaddress
\usepackage{amsmath}
\usepackage{graphicx,epstopdf}
\usepackage{gensymb}
\newcommand{\be}{\begin{equation}}
\newcommand{\ee}{\end{equation}}
\newcommand{\bea}{\begin{eqnarray}}
\newcommand{\eea}{\end{eqnarray}}
\newcommand{\bse}{\begin{subequations}}
\newcommand{\ese}{\end{subequations}}

\usepackage{color}
\usepackage[colorlinks,bookmarks=false,citecolor=darkblue,linkcolor=red,urlcolor=blue]{hyperref}

\definecolor{darkred}{rgb}{0.7,0.0,0.0}

\definecolor{darkblue}{rgb}{0,0.02,0.45}

\definecolor{darkgreen}{rgb}{0.02,0.45,0.0}

\definecolor{violet}{rgb}{0.8,0.2,0.6}

\begin{document}
	\title{Unconventional superparamagnetic behavior in the modified cubic spinel compound LiNi$_{0.5}$Mn$_{1.5}$O$_{4}$}
	\author{S. S. Islam}
	\author{Vikram Singh}
	\author{Somesh K}
	\author{Prashanta K Mukharjee}
	\affiliation{School of Physics, Indian Institute of Science Education and Research Thiruvananthapuram-695551, India}
	\author{A. Jain}
	\author{S. M. Yusuf}
	\affiliation{Solid State Physics Division, Bhabha Atomic Research Centre, Mumbai 400 085, India}
	\author{R. Nath}
	\email{rnath@iisertvm.ac.in}
	\affiliation{School of Physics, Indian Institute of Science Education and Research Thiruvananthapuram-695551, India}
	\date{\today}

\begin{abstract}
Structural, electronic, and magnetic properties of modified cubic spinel compound LiNi$_{0.5}$Mn$_{1.5}$O$_{4}$ are studied via x-ray diffraction, resistivity, DC and AC magnetization, heat capacity, neutron diffraction, $^7$Li nuclear magnetic resonance, magnetocaloric effect, magnetic relaxation, and magnetic memory effect experiments. We stabilized this compound in a cubic structure with space group $P4_{3}32$. It exhibits semiconducting character with an electronic band gap of $\Delta/k_{\rm B} \simeq 0.4$~eV. The interaction within each Mn$^{4+}$ and Ni$^{2+}$ sub-lattice and between Mn$^{4+}$ and Ni$^{2+}$ sublattices is found to be ferromagnetic (FM) and antiferromagnetic (AFM), respectively. This leads to the onset of a ferrimagnetic transition at $T_{\rm C} \simeq 125$~K. The reduced values of frustration parameter ($f$) and ordered moments reflect magnetic frustration due to competing FM and AFM interactions. From the $^7$Li NMR shift vs susceptibility plot, the average hyperfine coupling between $^7$Li nuclei and Ni$^{2+}$ and Mn$^{4+}$ spins is calculated to be $\sim 672.4$~Oe/$\mu_{\rm B}$. A detailed critical behaviour study is done in the vicinity of $T_{\rm C}$ using modified-Arrott plot, Kouvel-Fisher plot, and universal scaling of magnetization isotherms. The magnetic phase transition is found to be second order in nature and the estimated critical exponents correspond to the 3D XY universality class. A large magneto-caloric effect is observed with a maximum value of isothermal change in entropy $\Delta S_m \simeq - 11.3$~J/Kg-K and a maximum relative cooling power of $RCP \simeq 604$~J/Kg for 9~T magnetic field change. The imaginary part of the AC susceptibility depicts a strong frequency dependent hump at $T=T_{\rm f2}$ well below the blocking temperature $T_{\rm b}\simeq120$~K. The Arrhenius behaviour of frequency dependent $T_{\rm f2}$ and the absence of ZFC memory confirm the existence of superparamagnetism in the ferrimagnetically ordered state.
\end{abstract}
\maketitle

\section{Introduction}
\begin{figure*}
	\includegraphics[width=\linewidth]{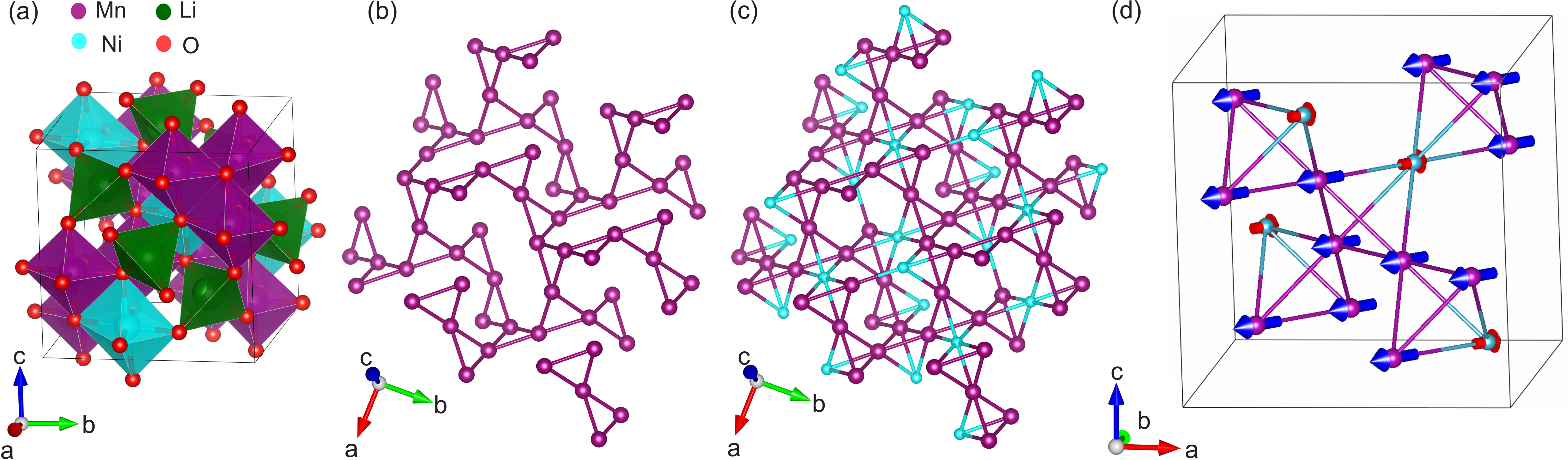}
	\caption{\label{Fig1} (a) Crystal structure of LNMO in three dimension. (b) Hyperkagome lattice made of Mn$^{4+}$ ions. (c) Pyrochlore lattice made of Ni$^{2+}$ and Mn$^{4+}$ ions. (d) A section of the pyrochlore lattice showing corner sharing tetrahedras and spin structure deduced from the neutron diffraction data at $T=5$~K.}
\end{figure*}
Geometrically frustrated quantum magnets are long been a field of attraction since they provide unique opportunity to realize novel quantum phases at low temperatures.\cite{Ramirez453,*Diep2013} One of the most studied geometrically frustrated systems in three dimension (3D) is the antiferromagnetic (AFM) pyrochlore lattice which features a network of corner sharing tetrahedras. Prominent examples in this category are compounds with general formula $AB_2$O$_4$ (spinels), $A_2B_2$O$_7$ (pyrochlores), and $AB_2$. In particular, in spinel oxides $AB_2$O$_4$, $B$ site ion forms a frustrated 3D pyrochlore lattice. Owing to ground state degeneracy, these compounds have witnessed various exotic low temperature phenomena ranging from quantum spin-liquid, spin-glass, field induced transitions, magnetization plateaus to heavy fermionic behaviour.\cite{Gardner53,Lee011004,Lacroix15178} In addition, there exists another series of compounds $AB_2$X$_6$, often referred as cubic modified pyrochlore lattice which mostly contains either mixed-valent or two kinds of transition metal ions. The compounds (Rb,Cs)Cr$_2$F$_6$,\cite{Ueda014701} (K,Rb)Os$_2$O$_6$,\cite{Bruhwiler020503} CsW$_2$O$_6$,\cite{Hirai166402} and CsNiCrF$_6$\cite{Fennell60} belong to this category and exhibit various exotic ground states.

Among the large class of spinel oxides, cubic LiMn$_2$O$_4$ (space group: $Fd\bar{3}m$, Mn$^{3+} \colon$ Mn$^{4+}=1\colon1$) is known to be a celebrated high-voltage cathode material for rechargeable Li-ion battery.\cite{Tarascon2859} It is reported to have charge ordering accompanied by orbital ordering due to the Jahn-Teller distortion in Mn$^{3+}$ ions and undergoes an antiferromagnetic (AFM) long-range-ordering (LRO) at low temperatures.\cite{Tomeno094422,*Sugiyama1187} In certain reports, the compound is found to show a spin-glass (SG) behaviour without any magnetic LRO which is attributed to the Mn$^{3+}$/Mn$^{4+}$ super-lattice charge order as well as the effect of frustration.\cite{Jang2504} These contradictory behaviours are believed to be originated from the Mn site disorder.\cite{Zhang104405} Recently, coexistence of LRO and SG state is found in LiMn$_2$O$_4$ nanorods.\cite{Zhang104405} The Ni doped LiNi$_{0.5}$Mn$_{1.5}$O$_{4}$ (abbreviated as LNMO) crystallizes in two different phases depending on the synthesis conditions. The stoichiometric LNMO has $P4_{3}32$ space group and exhibits a 1$\colon$3 cation order of Ni$^{2+}$ and Mn$^{4+}$ ions, while non-stoichiometric LiNi$_{0.5}$Mn$_{1.5}$O$_{4-\delta}$ has $Fd\bar{3}m$ space group.\cite{Kim906}
%In the $Fd-3m$ structure, Li atoms are located at the $8a$ site, O atoms at the $32e$ site, and Mn and Ni atoms are randomly distributed in the $16d$ site. On the other hand, for the $P_{4}332$ structure, Li atoms are located at the $8c$ site, O atoms at the $8c$ and $24e$ site, Ni atoms at the $4a$ site, and Mn atoms at the $12d$ site.

In the $Fd\bar{3}m$ structure, Ni and Mn atoms randomly occupy one crystallographic site while in the $P4_{3}32$ structure, they occupy two inequivalent sites independently.
%The Ni and Mn atoms are regularly ordered in different sites of the $P_{4}332$ structure while they are in the disordered state in the $Fd-3m$ structure.
In the $P4_{3}32$ structure which can also be referred as modified cubic spinel, the edge sharing of MnO$_6$ and NiO$_6$ octahedras and the corner shared LiO$_4$ tetrahedra lead to a complex three-dimensional (3D) structure [see Fig.~\ref{Fig1}(a)]. When only the interaction among the Mn$^{4+}$ ions is considered, LNMO forms a 3D network of corner sharing Mn$^{4+}$ triangles which is found to be a frustrated hyper-kagome lattice [see Fig.~\ref{Fig1} (b)]. Further, when the interaction between Mn$^{4+}$ and Ni$^{2+}$ ions are taken into account, a network of corner shared tetrahedras is formed where each tetrahedra consists of three Mn$^{4+}$ and one Ni$^{2+}$ ions. Hence, the hyper-kagome lattice transforms into a 3D pyrochlore lattice [see Fig.~\ref{Fig1}(c)]. Based on the preliminary magnetic measurements, LNMO is reported to show a magnetic transition at $T_{\rm C}\simeq 125$~K.\cite{Blasse383,*MukaiA672,*Amdouni100}
%A recent theoretical study suggested that the coupling within Ni and Mn sub-lattices is ferromagnetic (FM) while the coupling between two sub-lattices is antiferromagnetic (AFM).\cite{Xin128202}

In this paper, we present a detailed study of the physical properties of stoichiometric modified cubic spinel compound LNMO ($P4_{3}32$). A ferrimagnetic order is detected at $T_{\rm C} \simeq 125$~K. We found that the interaction within each Mn$^{4+}$ and Ni$^{2+}$ sub-lattice is ferromagnetic (FM), whereas the interaction between these two sublattices is antiferromagnetic (AFM), which results in a ferrimagnetic behaviour below $T_{\rm C}$. Multiple magnetic transitions are observed below $T_{\rm C}$, likely due to magnetic frustration. It exhibits magnetic relaxation and magnetic memory effect below $T_{\rm C}$, typically expected for superparamagnetic systems. A large magnetocaloric effect (MCE) is obtained across the magnetic transition. The critical analysis of magnetization and MCE data establish LNMO as a 3D XY type magnet. The paper is organized in the following manner. The experimental details concerning sample preparation and various measurements are described in Sec.~II. Section~III contains the experimental results which includes powder x-ray diffraction, resistivity, DC magnetization, heat capacity, neutron diffraction, $^{7}$Li NMR, magnetocaloric effect, AC susceptibility, magnetic relaxation, and magnetic memory effect measurements, followed by discussions. Our experimental findings are summarized in Sec.~IV.

\section{Methods}
Traditional sol-gel synthesis method was adopted to synthesize LNMO in polycrystalline form. At first, stoichiometric amount of lithium nitrate (LiNO$_{3}$, 99.99\%), manganese nitrate tetra-hydrate [Mn(NO$_3$)$_2$.4H$_2$O, $\geq 97$\%], and nickel nitrate hexa-hydrate [Ni(NO$_3$)$_2$.6H$_2$O, 99.999\%] were taken and dissolved into ethanol. The mixture was continuously stirred at 80~$^0$C until the whole solvent is evaporated from the mixture and dark black coloured paste was found. The resulting paste was then transferred into a crucible and preheated at 500~$^0$C for 2~hrs and then at 800$^0$C for 8~hrs. Subsequently, the furnace was switched off and the sample was cooled naturally within the furnace. The resultant sample was found to be formed in the space group $Fd\bar{3}m$, confirmed from the powder x-ray diffraction. In the next step, the resultant sample was ground thoroughly and pressed into pellets. The pellets were heated at 700~$^0$C for 2~days and then cooled very slowly to room temperature at a rate of 0.1$^0$C/min. This post firing of the $Fd\bar{3}m$ phase sample at 700~$^0$C was done to ensure the formation of the cation ordered $P4_{3}32$ phase. This method is well established and already experimented previously.\cite{Cai6908,*Lee3118}

Phase purity of the sample was checked from the high quality powder x-ray diffraction (XRD) data, collected using a PANalytical x-ray diffractometer (Cu K$_\alpha$ radiation, $\lambda_{\rm av} \simeq\ 1.5418$~\AA). The temperature dependent power x-ray diffraction was performed over a wide temperature range (15~K$\leq T \leq 300$~K). For going below room temperature, an Oxford Phenix low-temperature attachment to the diffractometer was used. To solve the magnetic structure, temperature dependent neutron powder diffraction (NPD) experiment was performed using the neutron powder diffractometer ($\lambda \simeq\ 1.094$~\AA) with three linear position-sensitive detectors at Dhruva reactor, Bhabha Atomic Research Center, India. Rietveld refinement of the powder XRD data and NPD data was performed using FullPROF software package.\cite{Rodriguez55}

The DC magnetization ($M$) was measured using a vibrating sample magnetometer (VSM) attachment to a commercial Physical Property Measurement System (PPMS, Quantum Design) as a function of temperature (2~K~$\leq T \leq$~600~K) and magnetic field (0 to 9~T). For the high temperature measurements ($T \geq 380$~K), a high-$T$ oven was attached to the VSM. Similarly, AC susceptibility was measured as a function of temperature (2~K~$\leq T \leq$~200~K) and frequency (50~Hz~$\leq \nu \leq$~10~kHz) in an AC field of 5~Oe using ACMS option of the PPMS. For the temperature dependent heat capacity ($C_{\rm p}$) measurement, the relaxation technique was adopted and the measurement was carried out on a pressed pellet using heat capacity option of the PPMS. Electrical resistivity ($\rho$) as a function of temperature was measured on a rectangular pellet using the four probe technique in PPMS.

The NMR measurements were performed by employing pulsed NMR technique on $^{7}$Li (nuclear spin $I=3/2$ and gyromagnetic ratio $\gamma_{N}/2\pi = 16.546$\,MHz/T) nuclei over a wide temperature range (4~K$\leq T \leq 290$~K). For this purpose, we have used a liquid helium cryostat (Janis, USA) with a field sweep superconducting magnet and a Tecmag (Redstone) spectrometer. The spectral measurements at different temperatures were carried out either by Fourier transform of the NMR echo signal at a fixed field of $H=1.5462$~T or by sweeping the field at a corresponding fixed frequency of 25.58~MHz. Traditional saturation recovery pulse sequence was used to measure the $^{7}$Li spin-lattice relaxation time ($T_1$).

\section{Results and Discussion}
\subsection{X-ray Diffraction}
\begin{figure}
	\includegraphics[width=\linewidth]{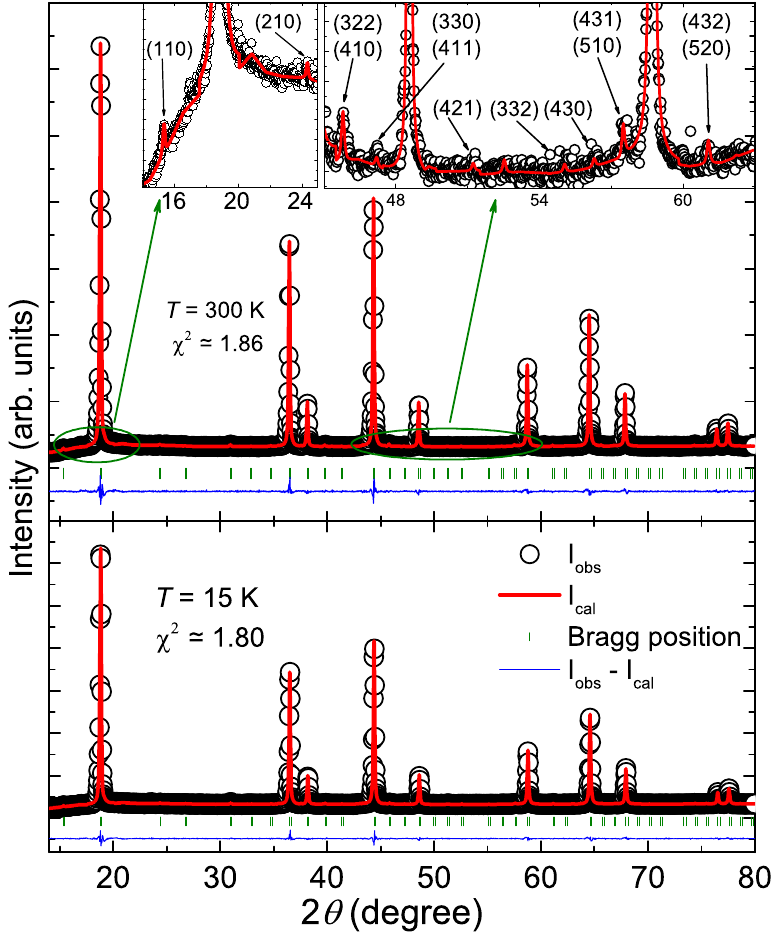}
	\caption{\label{Fig2} Upper panel: X-ray powder diffraction pattern (open circles) of LNMO at room temperature. The red solid line is the Rietveld refinement fit with $P4_3$32 space group. The expected Bragg positions are indicated by green vertical ticks. Blue solid line at the bottom denotes the difference between observed and calculated intensities. Inset: Enlarged portion of the XRD pattern to visualize small Bragg peaks. Lower panel: Rietveld refinement of the XRD pattern of LNMO at 15~K.}
\end{figure}

Figure~\ref{Fig2} presents the XRD pattern of LNMO at two end temperatures 300~K and 15~K. To evaluate the unit cell parameters and atomic positions, Rietveld refinement was performed on the powder XRD data. The initial structural parameters for this purpose were taken from Ref.~[\onlinecite{Cai6908,*Lee3118}]. All the peaks could be successfully indexed with cubic non-centrosymmetric space group $P4_3$32. The refined unit cell parameters, volume of the unit cell, and the atomic co-ordinates at room temperature are listed in Table~\ref{TableRefinement} which are in close agreement with the previous report. As already described in Sec.~I, LNMO exhibits 1:3 cation order, resulting in a superstructure cubic $P4_3$32 space group. The cation ordering in LNMO can be visualized by the emergence of several low angle and low intensity Bragg peaks, such as (110), (210), (322), (410) etc.\cite{Branford1649} These peaks are not allowed in the Bragg reflections in the normal face-centered cubic spinel ($Fd\bar{3}m$) structure, because of the reflection conditions $(h+k) = 2n$, $(h+l)= 2n$, and $(k+l) = 2n$. Here, $n$ is the integer and $(h,k,l)$ are the Miller indices. These low intensity peaks are highlighted in the inset of the upper panel of Fig.~\ref{Fig2}. It can be seen that all these small Bragg reflections could be perfectly indexed by using $P4_3$32 space group. We have also tried to do the refinement with $Fd\bar{3}m$ space group, which could not index these small Bragg peaks, thus confirming the phase purity of the sample with $P4_3$32 space group.

It is reported that the material synthesis following sol-gel method leads to the formation of nano-crystalline form of LNMO.\cite{KunduraciA1345,*Wang32} To estimate the crystallite size, we fitted the XRD peaks at room temperature by a Gaussian function and evaluated the full width at half maxima (FWHM) of the individual peak. Subsequently, by using the Scherrer equation, $\tau=K\lambda/\beta \rm{cos\theta}$ (where, $\tau$ is the crystallite size, $K$ is the dimensionless shape factor which has a typical value of 0.9, $\lambda$ is the x-ray wavelength, and $\beta$ is the line broadening at FWHM), the average crystallite size is calculated to be $\sim 60$~nm.\cite{Patterson978} Further, the analysis of the Scanning-Electron-Microscopy (SEM) data also reveals the nano-crystalline nature of LNMO sample with average particle size 100-150~nm.
\begin{figure}
	\includegraphics[width=\linewidth]{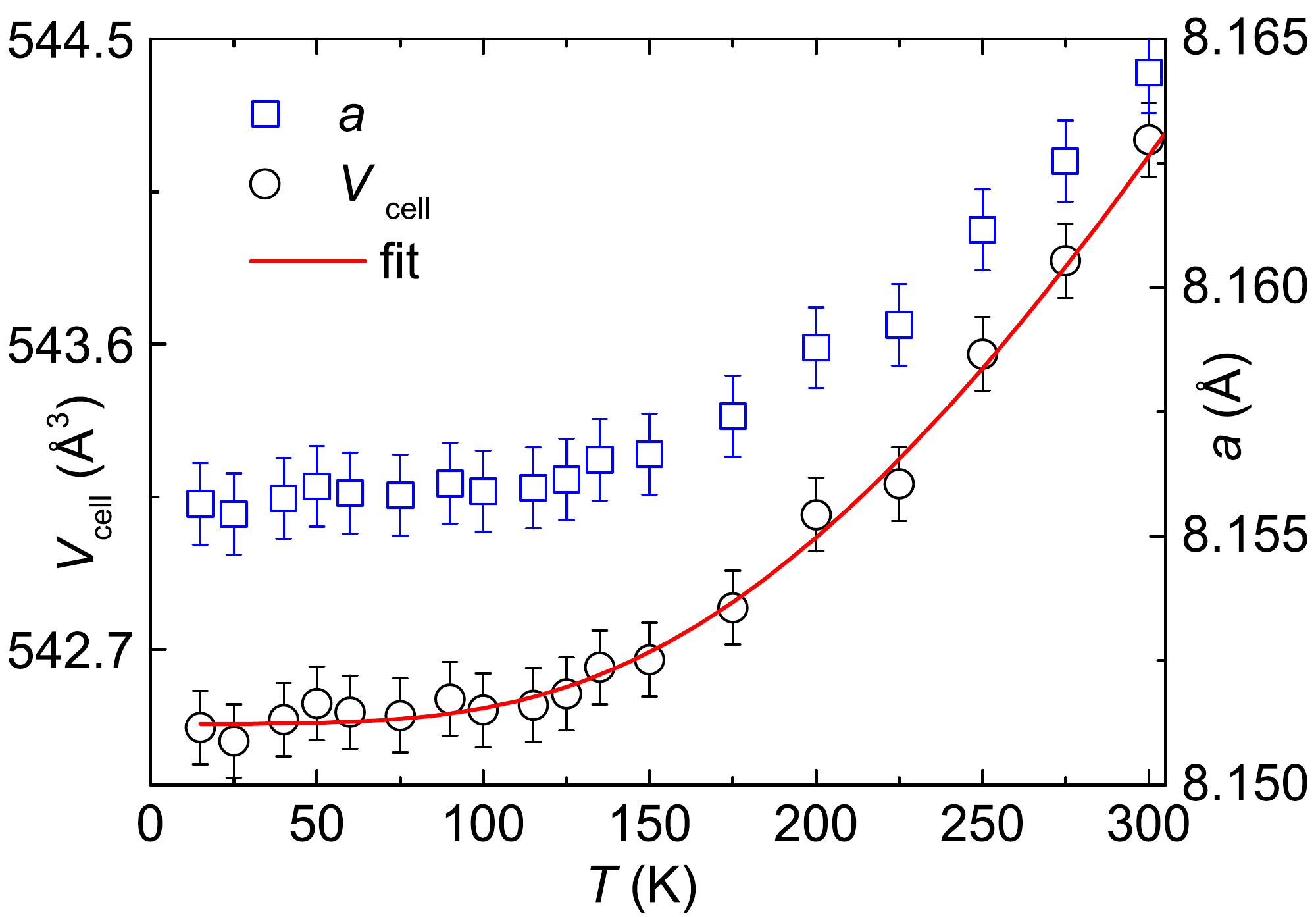}
	\caption{\label{Fig3} Variation of lattice constant ($a$) and unit cell volume ($V_{\rm cell}$) with temperature. The solid line represents the fit of $V_{\rm cell}(T)$ by Eq.~\eqref{VcellvsT}.}
\end{figure}
\begin{table}[ptb]
	\caption{Crystal structure data for LNMO at room temperature (cubic, space group: $P4_{3}32$). The obtained lattice parameters from the refinement are $a = b = c = 8.1643 (1)$~{\AA} and $ V_{\rm cell} \simeq 544.2~\AA^3 $. Our fit yields quality factors $R_{\rm p} \simeq 10.1$, $R_{\rm wp} \simeq 6.66$, $R_{\rm e} \simeq 4.88$, and goodness of fit $\chi^2 = [\frac{R_{\rm wp}}{R_{\rm e}}]^{2} \simeq 1.87$. Listed are the Wyckoff positions and the refined atomic coordinates ($x$, $y$, and $z$) for each atom.}
	\label{TableRefinement}
	\begin{ruledtabular}
		\begin{tabular}{ccccc}
			Atom & Site & $x$ & $y$ & $z$ \\\hline
			Li & $8c$ & $-0.0002 (4)$ & $-0.0002 (4)$ & $-0.0002 (4)$ \\
			Ni & $4b$ & $0.625$ & $0.625$ & $0.625$ \\
			Mn & $12d$ & $0.125$ & $0.3768 (3)$ & $0.8732 (3)$ \\
			O1 & $8c$ & $0.3816 (7)$ & $0.3816 (7)$ & $0.3816 (7)$ \\
			O2 & $24e$ & $0.1509 (6)$ & $-0.1395 (9)$ & $0.1238 (7)$ \\
		\end{tabular}
	\end{ruledtabular}
\end{table}

As shown in lower panel of Fig.~\ref{Fig2}, no extra peaks could be detected down to 15~K. Figure~\ref{Fig3} depicts the temperature variation of lattice constant ($a$) and unit cell volume [$V_{\rm cell}(T)$] obtained from the refinement. Both the quantities are found to decrease systematically during cooling and neither any structural transition nor any lattice distortion is observed in the entire measured temperature range (15~K$\leq T \leq 300$~K). Following the method described in Ref.~\cite{Islam174432}, $V_{\rm cell}(T)$ is fitted by
\begin{equation}
V_{\rm cell}(T)=\gamma U(T)/K_0+V_0,
\label{VcellvsT}
\end{equation}
where $V_0$ is the unit cell volume of the crystal structure at $T = 0$~K, $K_0$ is the bulk modulus, and $\gamma$ is the Gr$\ddot{\rm u}$neisen parameter. $U(T)$ is the internal energy and it can be expressed in terms of the Debye approximation as
\begin{equation}
U(T)=9p\,k_{\rm B}T\left(\frac{T}{\theta_{\rm D}}\right)^3\int_{0}^{\theta_{\rm D}/T}\dfrac{x^3}{e^x-1}dx.
\end{equation}
Here, $p$ is the number of atoms in the unit cell and $k_{\rm B}$ is the Boltzmann constant. The parameters evaluated from the fitting are Debye temperature $\theta_{\rm D} \simeq 965$~K, $\gamma/K_0 \simeq 1.24 \times 10^{-4}$~Pa$^{-1}$, and $V_0 \simeq 542.48$~\AA$^{3}$.

\subsection{Resistivity}
\begin{figure}
	\includegraphics[width=\linewidth] {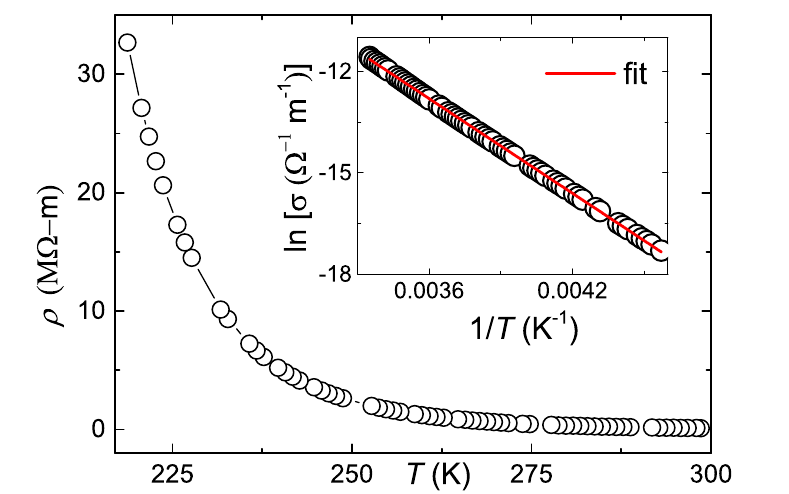}
	\caption{\label{Fig4} The electrical resistivity $\rho(T)$ of LNMO in zero field. Inset: ln($\sigma$) vs $1/T$. The solid line is the fit using Eq.~\eqref{res}.}
\end{figure}

Temperature dependent electrical resistivity [$\rho(T)$] measured in zero field is shown in Fig.~\ref{Fig4}. It increases rapidly with decreasing temperature which indicates that the ground state is insulating in nature. Below 218~K, $\rho(T)$ could not be measured since it exceeded the measurable range of the instrument. To evaluate the activation energy, the temperature dependent conductivity ($\sigma = 1/\rho$) data were fitted by Arrhenius equation
\begin{equation}
\sigma (T)= A exp \left(-\dfrac {\Delta}{k_{\rm B} T}\right),
\label{res}
\end{equation} 
where, $A$ is the proportionality constant and $\Delta$ is the activation energy. In the inset of Fig.~\ref{Fig4}, ln($\sigma$) is plotted against $1/T$ to highlight the activated behaviour. Our fit in the whole measured temperature range (218~K~$\leq$~$T$~$\leq$~300~K) yields $\Delta/k_{\rm B} \approx 0.4$~eV. This value of $\Delta/k_{\rm B}$ categorizes LNMO as a semiconductor. 
%The value of $\Delta/k_{\rm B}$ is also comparable with other semiconducting manganites such as LaMnO$_3$ ($\Delta/k_{\rm B}\simeq 0.215$~eV), HoMnO$_3$ ($\Delta/k_{\rm B}\simeq 0.373$~eV), NdMnO$_3$ ($\Delta/k_{\rm B}\simeq 0.245$~eV) etc.\cite{Martincarron527,*Jonker1424}

\subsection{DC Magnetization}
\begin{figure}
	\includegraphics[width=\linewidth] {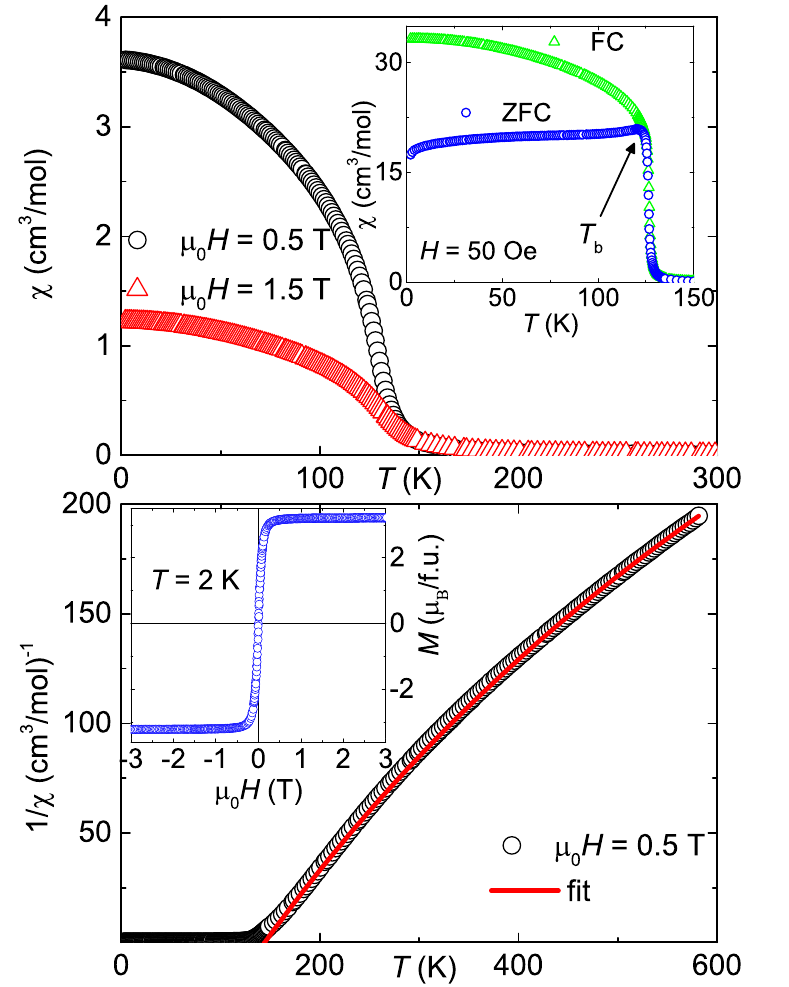}
	\caption{\label{Fig5} Upper panel: Temperature dependent DC magnetic susceptibility $\chi(T)$ at two different applied fields. Inset: $\chi(T)$ measured at 50~Oe in ZFC and FC conditions. Lower panel: $1/\chi$ vs $T$ in $\mu_{0}H = 0.5$~T. Inset: Magnetization isotherm at $T= 2$~K.}
\end{figure}
The upper panel of Fig.~\ref{Fig5} presents the temperature dependent DC magnetic susceptibility $\chi(T)$ ($\equiv M/H$) measured in an applied field of $H= 0.5$~T and 1.5~T. In the high temperature regime, $\chi(T)$ shows a gradual increase with decreasing temperature. Below about 140~K, $\chi(T)$ increases rapidly, indicating the onset of a ferrimagnetic/ferromagnetic ordering. From the $d\chi/dT$ vs $T$ plot, the ordering temperature is found to be $T_{\rm C}\simeq 125$~K. As depicted in the inset of the upper panel of Fig.~\ref{Fig5}, the zero-field cooled (ZFC) and field cooled (FC) susceptibility data at $H=50$~Oe show a significant bifurcation below $T_{\rm C}$. Such an irreversibility is a characteristic behaviour of ferrimagnetic/ferromagnetic compounds\cite{Nath224513} and is also observed for various SG\cite{Bag144436,Bag165977} and superparamagnetic\cite{Tsoi014445} systems. Moreover, the ZFC $\chi(T)$ shows a well defined maxima at $T_{\rm b}\simeq120$~K, which corresponds to the blocking temperature, typically expected for a superparamagnet.
A significant difference in the behaviour of FC $\chi(T)$ is expected between a superparamagnet and a SG system. For instance, FC magnetization always increases for a superparamagnet whereas for a SG system, it either remains flat or decreases with decreasing temperature, below $T_{\rm b}$.\cite{Sasaki104405,Bandyopadhyay214410}
As noticed from Fig.~\ref{Fig5}, the FC $\chi(T)$ below $T_{\rm b}$ increases monotonously with decreasing $T$, which is a primary indication of the superparamagnetic blocking process.\cite{Sasaki104405,Chen214436} In order to confirm this behaviour, we have performed a detailed ac susceptibility and magnetic memory effect experiments which are discussed later.
%As we have observed from the SEM experiment, our synthesis procedure indeed produces nanocrystalline form of LNMO.
%For samples in nano form, it is well-known that when the interparticle magnetic interaction is negligible or weak, the system may behave as a superparamagnet, while strong interparticle interactions lead to the formation of SG, if the system exhibits random and frustrated magnetic interactions.

The lower panel of Fig.~\ref{Fig5} shows the inverse magnetic susceptibility ($1/\chi$) for $\mu_{0}H=0.5$~T. In the paramagnetic regime ($T>T_{\rm C}$), $1/\chi$ typically shows a linear behaviour with temperature, due to uncorrelated moments. In contrast, the observed high temperature non-linear behaviour with a strong positive curvature is a possible signature of ferrimagnetic nature of LNMO.\cite{Nath224513,Kittel1976} To extract the magnetic parameters, we fitted the $\chi(T)$ data by the modified Curie-Weiss law
\begin{equation}
\chi(T)=\chi_0+\frac{C}{T-\theta_{\rm CW}}.
\label{cw}
\end{equation}
Here, $\chi_0$ is the temperature-independent susceptibility, $C$ is the Curie constant, and $\theta_{\rm CW}$ is Curie-Weiss temperature. Our fit in the high-temperature regime ($T \geq 450$~K) (see the lower panel of Fig.~\ref{Fig5}) yields the parameters: $\chi_{0} \simeq 0.0015 $~cm$^{3}$/mol, $C \simeq 1.604$~cm$^{3}$K/mol, and $\theta_{\rm CW} \simeq 144.4$~K. From the value of $C$, the effective moment is calculated to be $\mu_{\rm eff} = \sqrt{3k_{\rm B}C/N_{\rm A}\mu_{\rm B}^2} \simeq 3.58 \mu_{\rm B}$, where $N_{\rm A}$ is the Avogadro's number. This is close to the expected spin-only value of $\mu_{\rm eff} = \sqrt{\left (\dfrac{0.5\times[\mu_{\rm eff} (\rm Ni^{2+})]^2 + 1.5\times[\mu_{\rm eff} (\rm Mn^{4+})]^2}{0.5 + 1.5}\right)}~\mu_{\rm B} = 3.64~\mu_{\rm B}$, taking $\mu_{\rm eff} = 2.83$~$\mu_{\rm B}$ and 3.87~$\mu_{\rm B}$ for Ni$^{2+}$($S=1$) and Mn$^{4+}$($S=3/2$), respectively.\cite{MukaiA672} The positive value of $\theta_{\rm CW}$ implies that the dominant interaction among the magnetic ions is ferromagnetic (FM) in nature. The inset of the lower panel of Fig.~\ref{Fig5} shows a complete magnetization isotherm ($M$ vs $H$) at $T = 2$~K. It shows a very weak hysteresis and the magnetization saturates quickly at $\mu_{0}H = 0.5$~T, typically expected for a ferrimagnet. The saturation magnetization ($M_{\rm s}$) is found to be $\sim 3.2~\mu_{\rm B}$. Assuming a two sub-lattice model of magnetic species Ni$^{2+}$ and Mn$^{4+}$ with antiferromagnetic (AFM) coupling between them and using molecular-field approximation, the saturation magnetization for LNMO can be written as $M_{\rm s} = [S(\rm Mn^{4+}) \times g \times 1.5 - S(\rm Ni^{2+}) \times g \times 0.5]~\mu_{\rm B}$.\cite{Nath224513,Kittel1976} Taking $S(\rm Mn^{4+}) = 3/2$, $S(\rm Ni^{2+}) = 1$, and $g = 2$, $M_{\rm s}$ is calculated to be $3.5~\mu_{\rm B}$. Usually, the value of $g$ for Mn$^{4+}$ and Ni$^{2+}$ is always more than 2 which should produce $M_{\rm s}$ larger than $3.5~\mu_{\rm B}$.\cite{Sun343,*Werner214414,*Ruan55} Clearly, our experimental value of $M_{\rm s}$ is smaller than the expected spin-only value.

The extent of frustration in a spin system can be quantified by the frustration ratio $f=\frac{\vert\theta_{\rm CW}\vert}{T_{\rm C}}$.\cite{Ramirez453} According to the mean field theory $\theta_{\rm CW}$ is nothing but the sum of all exchange couplings present in the system i.e. $\theta_{\rm CW}=\sum_{i} J_{i}$, where $i$ is the number of nearest neighbor spins.\cite{Domb296} Typically, for a non-frustrated AFM system $\theta_{\rm CW} \sim T_{\rm C}$ (or $T_{\rm N}$) and $f$ has a value close to 1. However, for a highly frustrated antiferromagnet, $f$ value is much larger than 1 ($>10$).\cite{Ramirez423} On the other hand, for a system having FM and AFM interactions, the value of $\theta_{\rm CW}$ is reduced due to opposite sign of the exchange couplings. This results in a decreased value of $f$. For LNMO, the frustration ratio is calculated to be $f\simeq144.4/125\simeq1.16$. Since LNMO is having highly frustrated pyrochlore geometry, a reduced value of $f$ clearly implies co-existence of AFM and FM interactions in the system. Indeed, our neutron powder diffraction experiments (discussed later) confirm this proposition.

%LNMO comprises of two magnetic sublattices made of Mn$^{4+}$ and Ni$^{2+}$ ions. In each sublattice, magnetic ions are coupled ferromagnetically whereas the coupling between these two sublattices is antiferromagnetic.\cite{Xin128202} Thus, because of the co-existing AFM and FM interactions, the value of $f$ is reduced substantially despite a highly frustrated lattice geometry. 
%Ferromagnetic exchange interactions within the Ni and Mn sublattices and the AFM exchange interactions between the two sublattices could be reason for the ferrimagnetic order in LNMO.\cite{Xin128202} A previously reported neutron diffraction study on LNMO also supports the ferrimagnetic transition at $T_{\rm C}$.\cite{Gryffroy785}

\subsection{Heat Capacity}
\begin{figure}
	\includegraphics[width = \linewidth] {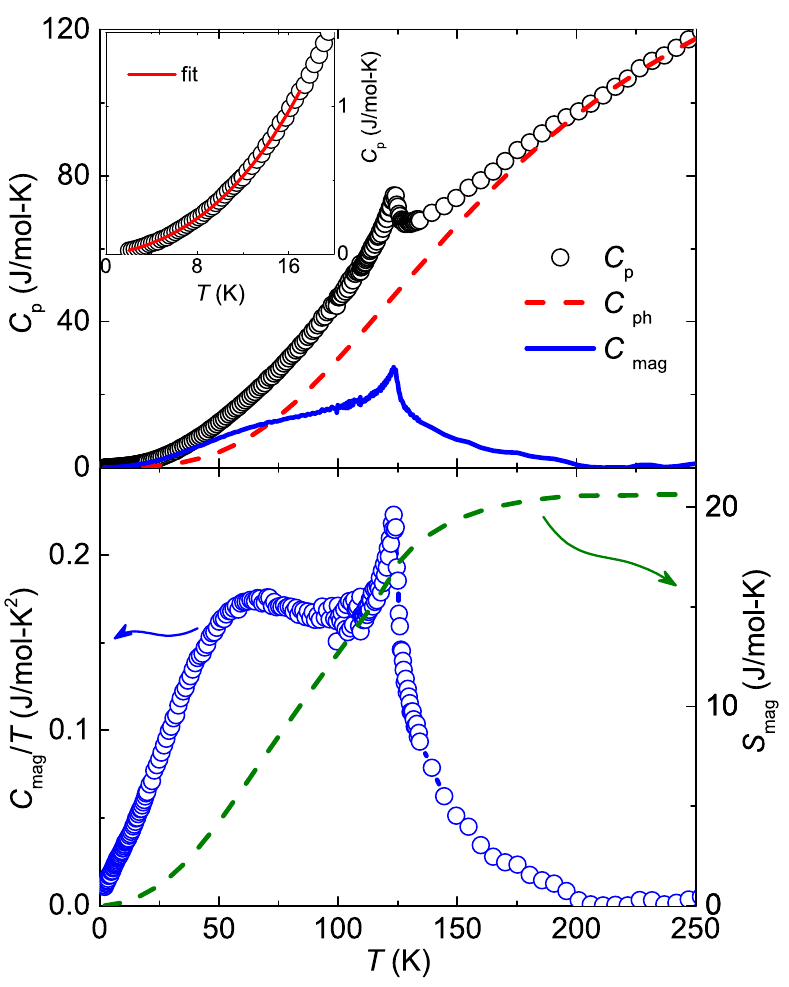}
	\caption{\label{Fig6} Upper panel: $C_{\rm p}(T)$ measured in zero magnetic field. The dashed and solid lines represent the phonon ($C_{\rm ph}$) and magnetic ($C_{\rm mag}$) contributions, respectively. Inset: Enlarged view of low temperature portion of $C_{\rm p}(T)$ data. The solid line is the fit as described in text.  Lower panel: $C_{\rm mag}/T$ and magnetic entropy ($S_{\rm mag}$) vs $T$ along the left and right $y$-axes, respectively.}
\end{figure}
The temperature dependent heat capacity [$C_{\rm p}(T)$] measured in zero field is shown in the upper panel of Fig.~\ref{Fig6}. A sharp and distinct peak is observed at $T_{\rm C} \simeq 124$~K, indicating the magnetic transition. In order to analyze the low temperature $C_{\rm p}(T)$ data, we first used the relation $C_{\rm p}(T) = \beta T^3$, which did not fit the data effectively. However, the low temperature $C_{\rm p}(T)$ data could be fitted nicely by adding an extra term $\delta T^{3/2}$ to the above relation, i.e., $C_{\rm p}(T) = \beta T^3 + \delta T^{3/2}$. Here, the first term ($\beta T^3$) represents the lattice contribution and the second term ($\delta T^{3/2}$) is typical for ferromagnetic/ferrimagnetic and glassy systems.\cite{Gopal2012,Thomson247} The inset of the upper panel of Fig.~\ref{Fig6} depicts the enlarged view of the low temperature portion of $C_{\rm p}(T)$. The solid line is the fit using the above relation, in the temperature range 2-16~K. The resulting $\beta$ and $\delta$ values are $\sim 1.07\times10^{-4}$~J~mol$^{-1}$ K$^{-4}$ and $\sim 0.0081$~J~mol$^{-1}$ K$^{-5/2}$, respectively. The electronic contribution $\gamma T$ is not considered in the fitting procedure, since LNMO is an insulator at low temperatures.

For an estimation of the phonon contribution $C_{\rm ph}(T)$, we fitted the experimental $C_{\rm p}(T)$ data in the high temperature regime ($T\geq 160~$K) by the Debye function
\begin{equation}
\label{Debye}
C_{\rm ph}(T) = 9R \left(\frac{T}{\theta_{\rm D}}\right)^3 \int_0^{\theta_{\rm D} / T} \frac{x^4e^x}{(e^x-1)^2} dx.
\end{equation}
Here, $R$ is the universal gas constant. The best fit was obtained with $\theta_{\rm D} \simeq 735$~K which is close to the value obtained from the $V_{\rm cell}(T)$ analysis. The high temperature $C_{\rm ph}(T)$ fit was extrapolated down to low temperatures and subtracted from the experimental $C_{\rm p}(T)$ data to obtain the magnetic contribution $C_{\rm mag}(T)$. The obtained $C_{\rm mag}(T)$ is shown as a solid line in the upper panel of Fig.~\ref{Fig6}. $C_{\rm mag}(T)/T$ vs $T$ is presented in the left $y$-axis of the lower panel of Fig.~\ref{Fig6}. The magnetic entropy $S_{\rm mag}(T)$ is calculated by integrating the $C_{\rm mag}(T)/T$ in the whole measured temperature range as
\begin{equation}
\label{entropy}
S_{\rm mag}(T) = \int_{2.1\rm K}^{T} \frac{C_{\rm mag}(T)}{T} dT.
\end{equation}
As shown in the lower panel of Fig.~\ref{Fig6}, the value of $S_{\rm mag}$ is found to be $\sim 20.6$~J/mol-K at 250~K, which is very close to the theoretically expected value of $S_{\rm mag} = 0.5 \times R ln[2 S (\rm Ni^{2+})+1]$~+~$1.5 \times R ln[2 S(\rm Mn^{4+})+1] = 21.8$~J/mol-K. One broad hump is observed in $C_{\rm mag}(T)$ at around $T\sim 50$~K which cannot be attributed to any magnetic transition as $\chi(T)$ does not show any feature at this temperature. Similar feature is reported earlier for BaMn$_2$As$_2$ and BiMn$_{2}$PO$_{6}$ where it is proposed that this broad hump is associated with the temperature dependent change in the population of the Zeeman levels below transition, arising due to the temperature dependent exchange field.\cite{Johnston094445,Nath024431} Using a Weiss molecular field theory on a Heisenberg model, it is predicted that the hump like feature is pronounced for the systems with higher spin values.\cite{Johnston094445,Nath024431} Thus, the observed hump in $C_{\rm mag}(T)$ is obvious since LNMO has two high spins: $S=1$ and $S=3/2$.

\subsection{Neutron Diffraction}
\begin{figure}
	\includegraphics[width=\linewidth] {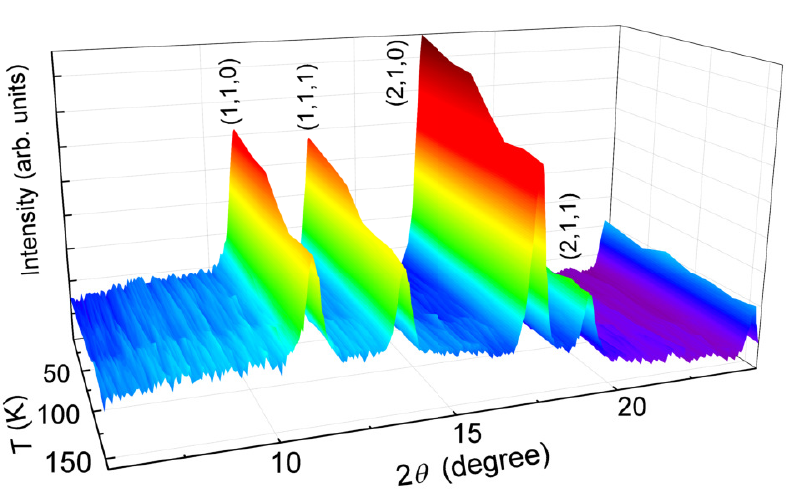}
	\caption{\label{Fig7} Temperature evolution of the neutron powder diffraction pattern of LNMO shown for the low angle regime. An enhancement in the intensity of fundamental nuclear Bragg peaks is evident below $T_{\rm C}$. These Bragg peaks at $2\theta \simeq 10.81\degree$, $13.31\degree$, $17.17\degree$, and $18.84\degree$ correspond to (1,1,0), (1,1,1), (2,1,0), and (2,1,1) planes, respectively.}
\end{figure}
\begin{figure}
	\includegraphics[width=\linewidth] {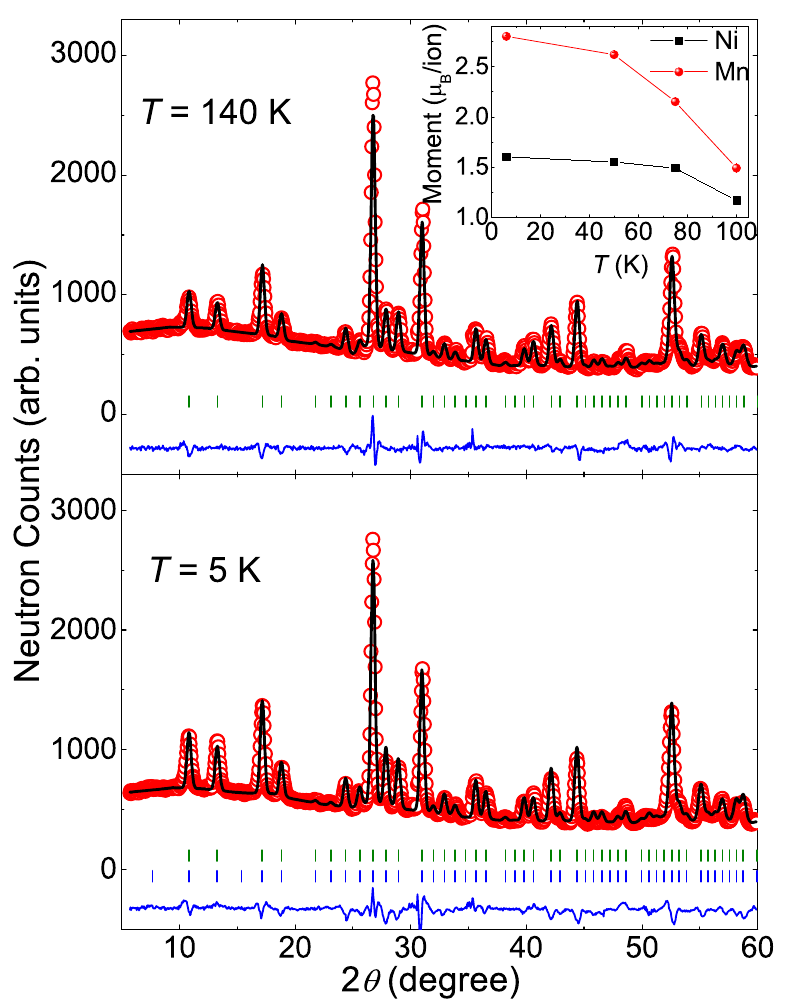}
	\caption{\label{Fig8} Rietveld refined neutron powder diffraction patterns at $T = 140$~K (upper panel) and $T = 5$~K (lower panel). Open circles represent the experimental data, solid line represents the calculated curve, and difference between them is shown as a solid line at the bottom. Vertical marks correspond to the position of all allowed Bragg reflections for the nuclear (top row) and magnetic (bottom row) reflections. Inset: Temperature variation of ordered magnetic moment for Ni$^{2+}$ and Mn$^{4+}$ ions.}
\end{figure}

Figure~\ref{Fig7} depicts a series of neutron powder diffraction (NPD) patterns for LNMO, over the temperature range $5 - 300$~K. An enhancement in the intensity of fundamental nuclear Bragg peaks at $\sim 10.81\degree$, $13.31\degree$, $17.17\degree$, and $18.84\degree$ are observed below $T_{\rm C}$. The observation of additional intensity at the positions of fundamental nuclear Bragg reflections (with no additional peaks) indicate the presence of a ferro- or ferrimagnetic ordering. As shown in the upper panel of Fig.~\ref{Fig8}, all the nuclear peaks at $T=140$~K could be refined using cubic crystal structure with space group $P4_{3}32$. The refined structural parameters are listed in Table~\ref{neutron}. These values are in close agreement with the refined values obtained from the powder XRD data. The lower panel of Fig.~\ref{Fig8} shows the Rietveld refinement of NPD pattern at $T = 5$~K. All the low angle peaks with additional intensity could be indexed with a propagation vector $k = (0, 0, 0)$ and space group $P4_{3}32$. The symmetry analysis shows that the observed NPD patterns can be fitted assuming a collinear ferrimagnetic structure with magnetic moments aligned along the [110] direction.
\begin{table}[ptb]
	\caption{Structural parameters of LNMO refined from the neutron diffraction data at $T = 140$~K (Structure: cubic and Space group: $P4_{3}32$). The obtained lattice parameters from the refinement are $a = b = c = 8.2000(8)$~\AA and $ V_{\rm cell} \simeq 551.36(9)~\AA^3 $. The fit yields quality factors $R_{\rm p} \simeq 21.5$, $R_{\rm wp} \simeq 18.2$, $R_{\rm e} \simeq 15.8$, and goodness of fit $\chi^2 = [\frac{R_{\rm wp}}{R_{\rm e}}]^{2} \simeq 1.327$. Listed are the Wyckoff positions and the refined atomic coordinates ($x$, $y$, and $z$) for each atom.}
	\label{neutron}
	\begin{ruledtabular}
		\begin{tabular}{ccccc}
			Atom & Site & $x$ & $y$ & $z$ \\\hline
			Li & $8c$ & $0.0002(19)$ & $0.0002(19)$ & $0.0002(19)$ \\
			Ni & $4b$ & $0.625$ & $0.625$ & $0.625$ \\
			Mn & $12d$ & $0.125$ & $0.3787(11)$ & $0.8713(11)$ \\
			O1 & $8c$ & $0.3809(5)$ & $0.3809(5)$ & $0.3809(5)$ \\
			O2 & $24e$ & $0.1503(4)$ & $-0.1417(5)$ & $0.1298(5)$ \\
		\end{tabular}
	\end{ruledtabular}
\end{table} 
The magnetic spin structure determined from Rietveld refinement of the NPD pattern at $T = 5$~K is shown in Fig.~\ref{Fig1}(d). Within each Ni$^{2+}$ or Mn$^{4+}$ sublattice, the moments are arranged parallel to each other providing a FM intra-sublattice interaction whereas between the Ni$^{2+}$ and Mn$^{4+}$ sublattices the alignment is found to be antiparallel which provides an AFM inter-sublattice interaction. From Fig.~\ref{Fig1}(d) it is clearly evident that the system is still frustrated due to competing FM and AFM interactions. However, the extent of frustration is definitely less than a conventional AFM pyrochlore lattice. This indeed fall in line with the $\chi(T)$ analysis where the reduced value of $f$ is attributed to the co-existence of AFM and FM interactions.

One can also understand the exchange interactions by looking at the bond angles. From the refinement of NPD data at $T = 5$~K, we found the angles $\angle Mn - O1 - Mn \simeq 93.8$~$\degree$, $\angle Mn - O2 - Mn \simeq 98.4$~$\degree$, and $\angle Mn - O2 - Ni \simeq 96.80$~$\degree$ and $95.3$~$\degree$. According to Goodenough-Kanamori rule,\cite{Kanamori87,*Goodenough564} the superexchange interaction between Mn$^{4+}$ ($3d^3$) ions through the O$^{2-}$ ($2p^2$) ion is expected to be AFM when the angle $\angle Mn - O - Mn$ is linear ($\sim 180^{\degree}$) and it crosses over to FM interaction for the angle $< 135^{\degree}$.\cite{Shimakawa1249} On the contrary, for the interaction between Mn$^{4+}$ ($3d^3$) and Ni$^{2+}$ ($3d^8$) ions via O$^{2-}$ ($2p^2$) ion, it is reported that FM interaction occurs for angle $\angle Mn - O - Ni$ close to $180^{\degree}$ and AFM interaction for the angle close to $90^{\degree}$.\cite{Kanamori87} Thus, the obtained FM interaction between Mn$^{4+}$ ions and AFM interaction between Mn$^{4+}$ and Ni$^{2+}$ ions are consistent with the bond angle analysis using NPD data.
 
%Indeed, similar spin structure with planar symmetry has been observed in several compounds in the field induced 3D-XY ordered state such as (Hpip)$_2$CuBr$_4$ and TlCuCl$_3$.\cite{Thielemann020408,Tanaka939}
%XY-type ordered state has also been visualized via neutron diffraction in high $T_{\rm C}$ superconductor cuprates such as in La$2$CuO$_4$ and in some pyrochlore aniferromagnets such as Er$_2$Ti$_2$O$_7$ and Yb$_2$Ti$_2$O$_7$.\cite{Champion020401,Hallas105}

At $T = 5$~K, the refined values of the ordered moment at $4b$ (Ni$^{2+}$) and $12d$ (Mn$^{4+}$) sites are found to be $\mu = 1.60(1)$~$\mu_{\rm B}$ and 2.80(1)~$\mu_{\rm B}$, respectively, which are smaller as compared to the expected spin only values (2~$\mu_{\rm B} /\rm {Ni^{2+}}$ for $S=1$ and 3~$\mu_{\rm B} /\rm{Mn^{4+}}$ for $S = 3/2$), assuming $g=2$. Such a reduced moment is typically observed in low-dimensional and frustrated spin systems which is attributed to the effect of quantum fluctuations and magnetic frustration, respectively.\cite{Manna224420} As shown in the inset of Fig.~\ref{Fig8} the value of the ordered moment for both Ni$^{2+}$ and Mn$^{4+}$ decreases as the temperature approaches $T_{\rm C}$, typical behaviour of sub-lattice magnetization in the ordered state. Using these values of the ordered moments at $T = 5 $~K, one expects a saturation magnetization of $M_{\rm s} \simeq (\mu_{\rm Mn^{4+}} \times 1.5 - \mu_{\rm Ni^{2+}} \times 0.5)~\mu_{\rm B} \simeq 3.4$~$\mu_{\rm B}$, assuming a ferrimagnetic spin structure which is indeed consistent with $M_{\rm s} \simeq 3.2$~$\mu_{\rm B}$ obtained from the $M$ vs $H$ curve at $T = 2$~K.
%It is interesting to note that the spin alignments are restricted only to the $ab$-plane [Fig.~\ref{Fig1}(d)] which possibly indicates a XY-type anisotropy.

\subsection{$^7$Li NMR}
Since Li is coupled strongly with the magnetic Mn$^{4+}$ and Ni$^{2+}$ ions, (see the inset of Fig.~\ref{Fig9}) it is possible to get the information about the static and dynamic properties of the spins by performing $^7$Li NMR.
\subsubsection{$^7$Li NMR Spectra}
\begin{figure}
	\includegraphics[width = \linewidth] {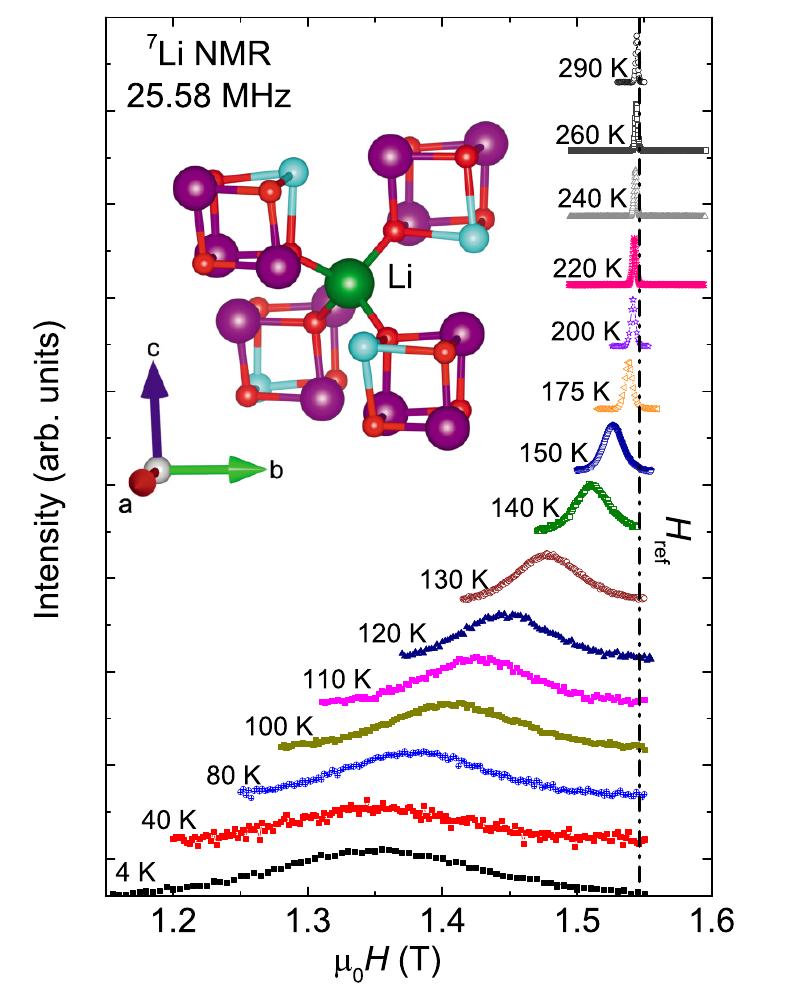}
	\caption{\label{Fig9} Field-sweep $^{7}$Li NMR spectra at different temperatures for the polycrystalline LNMO sample measured at 25.58\,MHz. The vertical dashed line corresponds to the $^{7}$Li resonance frequency of the non-magnetic reference. Inset: Coupling of Li nucleus with four cubic units made of Ni$^{2+}$ and Mn$^{4+}$ ions.}
\end{figure}
\begin{figure}
	\includegraphics[width=\linewidth] {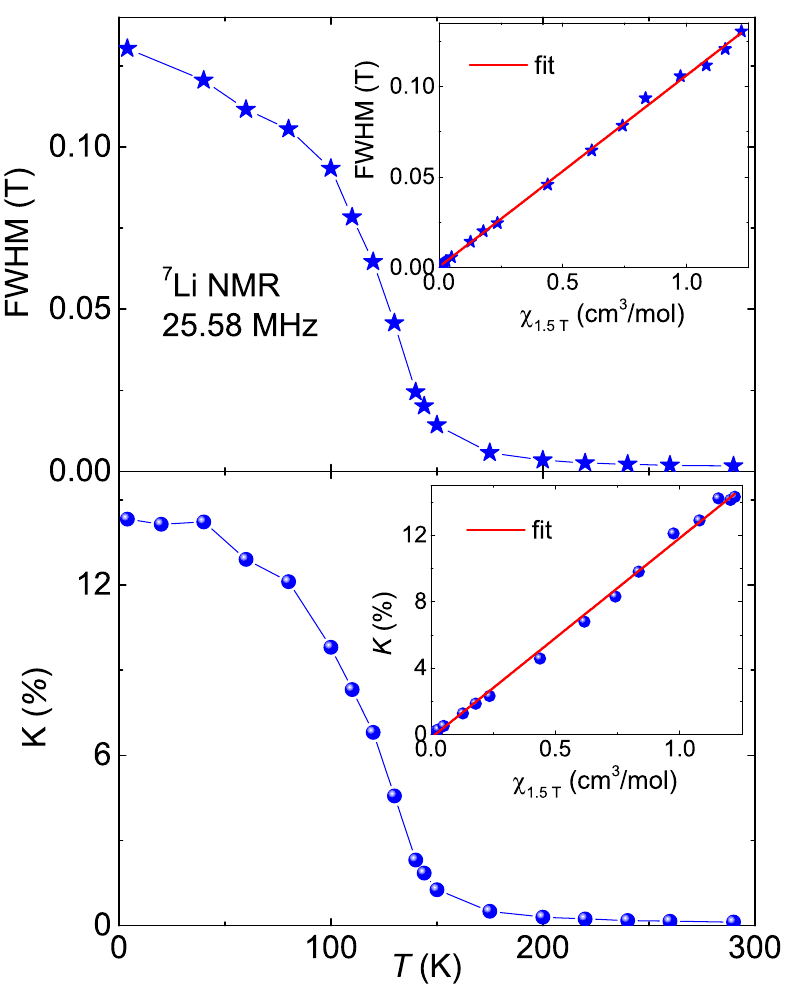}
	\caption{\label{Fig10} Upper panel: Fullwidth at half maximum (FWHM) of $^7$Li NMR spectra plotted as a function temperature. Inset: FWHM vs $\chi$ measured at 1.5~T is plotted with temperature as an implicit parameter. The solid line is a linear fit. Lower panel: Temperature-dependent $^{7}$Li NMR shift ($K$) vs temperature. Inset: $^{7}$Li NMR shift vs $\chi$ measured at 1.5~T is plotted with temperature as an implicit parameter. The solid line is a linear fit.}
\end{figure}
For a quadrupolar $^7$Li ($I = 3/2$) nucleus, one would expect two satellite peaks along with the central line due to three allowed transitions. Our $^7$Li NMR spectra, however, display a single spectral line in the whole temperature range as shown in Fig.~$\ref{Fig9}$ which corresponds to the central transition ($+1/2 \leftrightarrow -1/2$). The absence of satellite peaks could be either due to low quadrupolar frequency or the distribution of the satellite peak intensity over a wide frequency/field range. A single spectral line in $^7$Li NMR is typically observed in low-dimensional oxides.\cite{Ranjith014415,*Ranjith024422} The line shape is found to be symmetric in the entire measured temperature range suggesting the absence of magnetic anisotropy in the compound. The NMR line broadens drastically below $T_{\rm C}$ which reflects that Li nucleus senses the static internal field in the ordered state. With decreasing temperature, the peak position of the spectral line is found to be shifted. The upper panel of Fig.~$\ref{Fig10}$ depicts the fullwidth at half maximum (FWHM) of $^7$Li NMR spectra plotted as a function temperature. At high temperatures ($T\geq 200$~K), it is almost temperature independent, increases abruptly below about 150~K (near $T_{\rm C}$), and then levels off to a constant value with lowering temperature. The over all temperature dependent behaviour of FWHM is similar to that of the bulk $\chi(T)$ data. The FWHM is plotted against $\chi$ with temperature as an implicit parameter in the inset of the upper panel of Fig.~\ref{Fig10}. It indeed gives a straight line behaviour over the whole measured temperature range implying that the linewidth traces the bulk $\chi(T)$.

The $^7$Li NMR shift ($K$) was extracted from the central peak positions of the spectra in Fig.~$\ref{Fig9}$ by using the relation $K(T) =[H_{\rm ref} - H(T)]/H(T)$, where $H(T)$ and $H_{\rm ref}$ are the resonance field of the sample and the nonmagnetic reference sample, respectively. The temperature dependent NMR shift [$K(T)$] is presented in the lower panel of Fig.~$\ref{Fig10}$. It shows a constant behaviour at high temperatures and then increases abruptly below 150~K suggesting the occurrence of a ferrimagnetic ordering. The overall behaviour of $K(T)$ is again similar to the that observed for $\chi(T)$ and FWHM($T$). Since $K(T)$ measures the intrinsic spin susceptibility $\chi_{\rm spin}(T)$, one can write the relation
\begin{equation}
K(T)=K_{\rm chem}+\frac{A_{\rm hf}}{N_{\rm A}\mu_{\rm B}} \chi_{\rm spin}(T),
\label{shift}
\end{equation}
where $K_{\rm chem}$ is the temperature-independent chemical shift and $A_{\rm hf}$ is the hyperfine coupling constant between the Li nuclei and electronic (Ni$^{2+}$, Mn$^{4+}$) spins. The $K$ vs $\chi$ plot with temperature as an implicit parameter is presented in the inset of the lower panel of Fig.~$\ref{Fig10}$. It produces a linear behaviour in the entire measured temperature range. From the slope of a straight line fit, the transfer hyperfine coupling is evaluated to be $A_{\rm hf}\simeq 672.4$~Oe/$\mu_{\rm B}$.

\subsubsection{Spin-lattice relaxation rate 1/$T_1$ }
\begin{figure}
	\includegraphics[width=\linewidth] {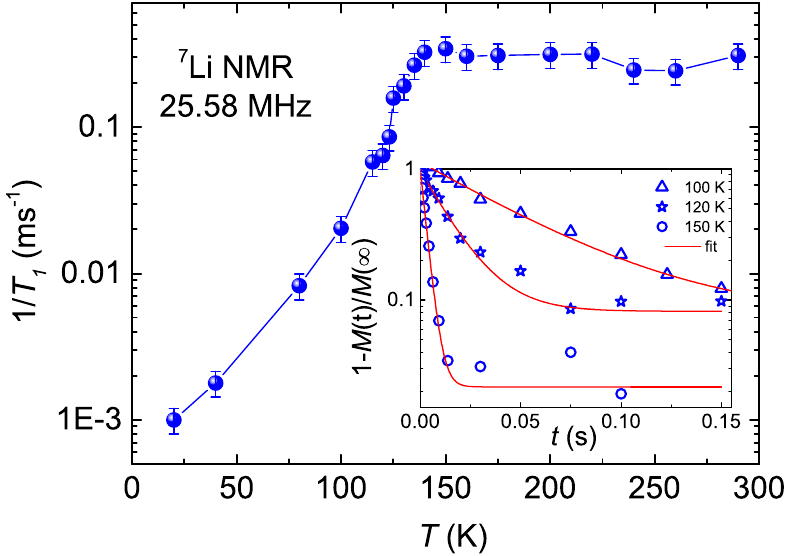}
	\caption{\label{Fig11} Spin-lattice relaxation rate $1/T_{1}$ vs temperature measured at 25.58\,MHz. Inset: The longitudinal magnetization recovery curves at various temperatures. The solid lines are the fits, as described in the text.}
\end{figure}
In order to probe the low energy spin dynamics, $^7$Li NMR spin-lattice relaxation rate ($1/T_1$) is measured at the field corresponding to the central peak position at different temperatures. The measurements are carried out in the temperature range 4~K to 290~K. Recovery of the longitudinal magnetization is monitored after a saturation pulse sequence. Some of the representative recovery curves in the low temperature regime are shown in the inset of Fig.~$\ref{Fig11}$. These recovery curves are fitted by single exponential function\cite{Ranjith014415}
\begin{equation}
1-\frac{M(t)}{M_{0}}=Ae^{-t/T_{1}},
\label{recovery}
\end{equation}
where $M(t)$ is the nuclear magnetization at a time $t$ after the saturation pulse and $M_{0}$ is the equilibrium magnetization. The $1/T_1$ extracted from the fit is plotted with respect to the temperature in Fig.~$\ref{Fig11}$.

At high temperatures ($T \geq 160$~K), $1/T_1$ is almost temperature independent. In the paramagnetic regime, where the temperature is higher than the exchange energy between the spins, the temperature independent behaviour of $1/T_1$ is obvious due to uncorrelated moments.\cite{Moriya23} Typically, when one approaches the magnetic transition from high temperatures, $1/T_1$ is anticipated to show a sharp peak or divergence with temperature due to the critical slowing down of the fluctuating moments. Our $1/T_1$ does show a weak anomaly at $T_{\rm C}$ and then decreases rapidly below $T_{\rm C}$. The decrease below $T_{\rm C}$ indicates the relaxation
due to scattering of magnons by the nuclear spins.\cite{Ranjith014415}
%Similar behavior in $1/T_1(T)$ has also been reported earlier in few other ferrimagnets such as Cu$_2$OSeO$_3$ and [MnTFPP][TCNE].\cite{Belesi094422,*Fardis064422}
%On the contrary, our system do not show any peak at $T_{\rm C}$. It rather shows a rapid decease with temperature, below $T_{\rm C}$ which can be explained by the partial cancellation of the fluctuations at the Li site. As shown in inset of Fig.~$\ref{Fig9}$, each Li atom in the crystal structure sits in a symmetric position with respect to four surrounding cubic units. Each cubic unit is made up of three Mn atoms, one Ni atom, and three oxygen atoms which are situated at the corners of each cube. Because of the 3D symmetric arrangement of the Mn$^{4+}$ and Ni$^{2+}$ ions around the Li ion, the spin fluctuations at the Li site get partially filtered out when the correlations become stronger in the ordered state. Therefore, instead of a peak/divergence, $1/T_1$ at the $^7$Li site shows a gradual decrease below $T_{\rm C}$.

\subsection{Critical Scaling of Magnetization}
\begin{figure*}
	\includegraphics[width= \linewidth] {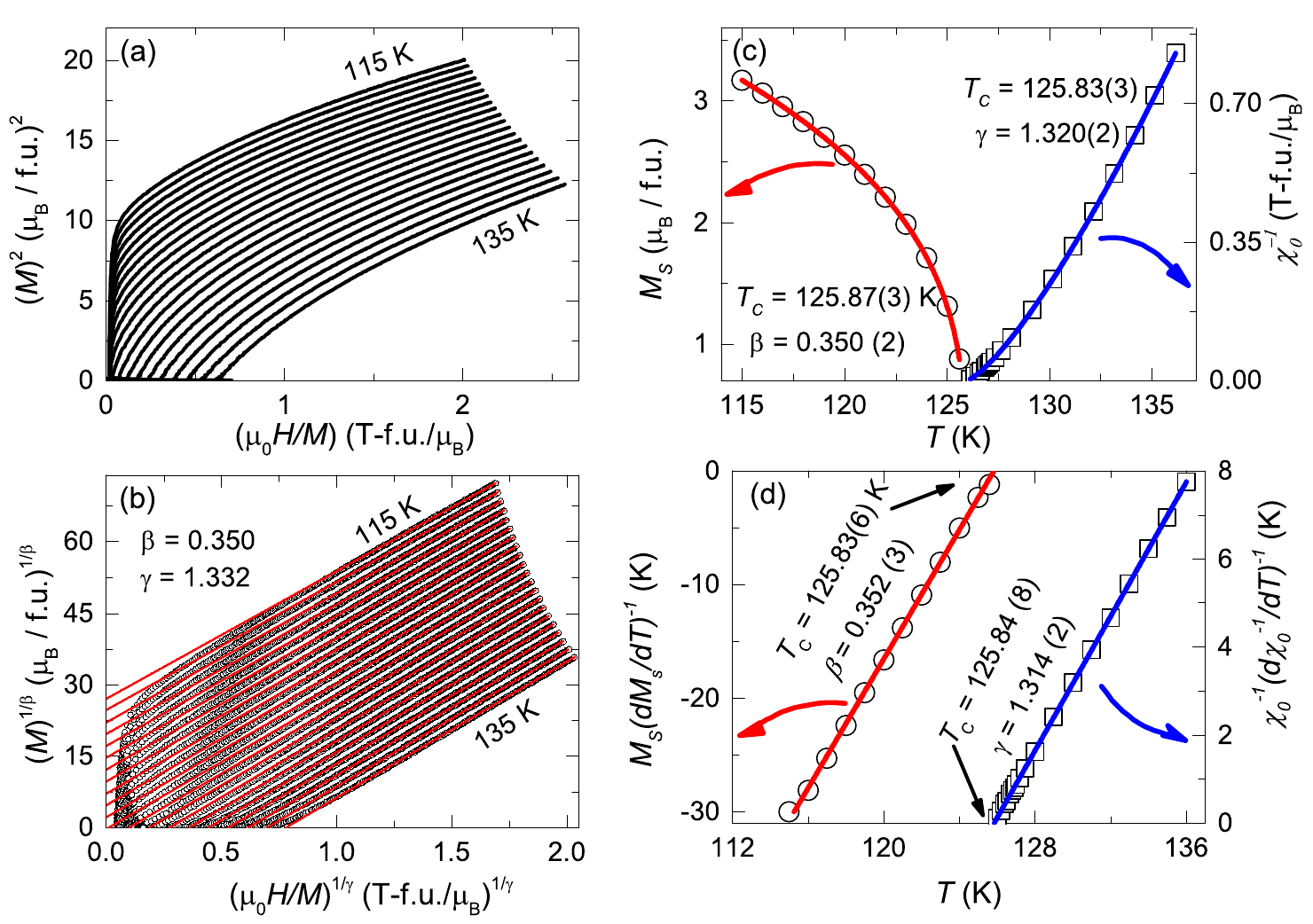}
	\caption{\label{Fig12}(a) The Arrott plots ($M^2$ vs $H/M$) for LNMO at different temperatures, above and below $T_{\rm C}$. (b) The modified Arrott plots ($M^{1/\beta}$ vs $(H/M)^{1/\gamma}$) for LNMO at different temperatures. The solid lines are the linear fits to the data in the high field regime ($H\geq~2.5$~T) and are extrapolated to $H/M=0$. (c) Spontaneous magnetization $M_{\rm S}$ and zero field inverse susceptibility  $\chi_0^{-1}$ as a function of temperature in the left and right $y$-axes, respectively, obtained from the intercepts of the modified Arrott plots in the vicinity of $T_{\rm C}$. The solid lines are the fits, as described in the text. (d) The Kouvel-Fisher plots for $M_{\rm S}$ and $\chi_0^{-1}$. The solid lines are the linear fits.}
\end{figure*}
For a better understanding of the magnetic properties, we have performed the critical behaviour study of magnetization in the vicinity of critical temperature $T_{\rm C}$. Scaling hypothesis suggests that the second order phase transition near $T_{\rm C}$ can be characterized by a set of critical exponents ($\beta$, $\gamma$, and $\delta$) and magnetic equations of state.\cite{stanley1971} Spontaneous magnetization $M_{\rm S}(T)$ at $T<T_{\rm C}$, zero field inverse susceptibility $\chi_0^{-1}(T)$ at $T>T_{\rm C}$, and isothermal magnetization ($M$ vs $H$) at $T=T_{\rm C}$ are related to the critical exponents by the following equations:
\begin{equation}
M_S(T)=M_0(-\varepsilon)^\beta, \varepsilon<0,
\label{cre1}
\end{equation}
\begin{equation}
\chi_0^{-1}(T)=\varGamma(\varepsilon)^\gamma, \varepsilon>0,
\label{cre2}
\end{equation}
\begin{equation}
M= X H^{1/\delta}, \varepsilon=0,
\label{cre3}
\end{equation}
where $\varepsilon= (T-T_{\rm C})/T_{\rm C}$ is the reduced temperature and $M_0$, $\varGamma$, and $X$ are the critical amplitudes.

Usually, the Arrott-Noakes equation of state\cite{Arrott786} is employed for a reliable estimation of critical exponents and $T_{\rm C}$ using magnetic isotherms, which can be written as
\begin{equation}
(H/M)^{1/\gamma}=A\varepsilon+BM^{1/\beta}.
\label{AN}
\end{equation}
According to this equation, for a particular set of $\beta$ and $\gamma$ values, $M^{1/\beta}$ vs $(H/M)^{1/\gamma}$ plots should produce a set of parallel straight lines in the high field region for different temperatures around $T_{\rm C}$ and the isotherm at $T=T_{\rm C}$ should pass through origin. The plot of $M^{1/\beta}$ vs $(H/M)^{1/\gamma}$ is often called the modified Arrott plot (MAP) and the exponents reflect the universality class of the spin system.
The mean field exponents ($\beta = 0.5$ and $\gamma = 1$) lead to conventional Arrott plot ($H/M$ vs $M^2$), which is traditionally used for critical behaviour analysis.\cite{Arrott1394} We have measured several magnetic isotherms around the $T_{\rm C}$ in close temperature steps. To avoid residual magnetization, at each temperature the measurements are done after cooling the sample in zero field from high temperatures (above $T_{\rm C}$). Figure~\ref{Fig12}(a) presents the standard Arrott plots for LNMO. All the curves show a non-linear behaviour and a downward curvature in the high field region. This indicates that the transition is non-mean field type and standard Arrott plot (associated with mean field theory) may not be an appropriate way to analyze the critical behaviour of this system. Further, the order of phase transition can be determined from the slope of Arrott plots. According to the Banerjee criterion, the positive slope corresponds to a second order phase transition whereas the negative slope indicates a first order phase transition.\cite{Banerjee16} Thus, the positive slope found in Fig.~\ref{Fig12}(a) confirms second order nature of the transition.

\begin{figure}
	\includegraphics[width=\linewidth] {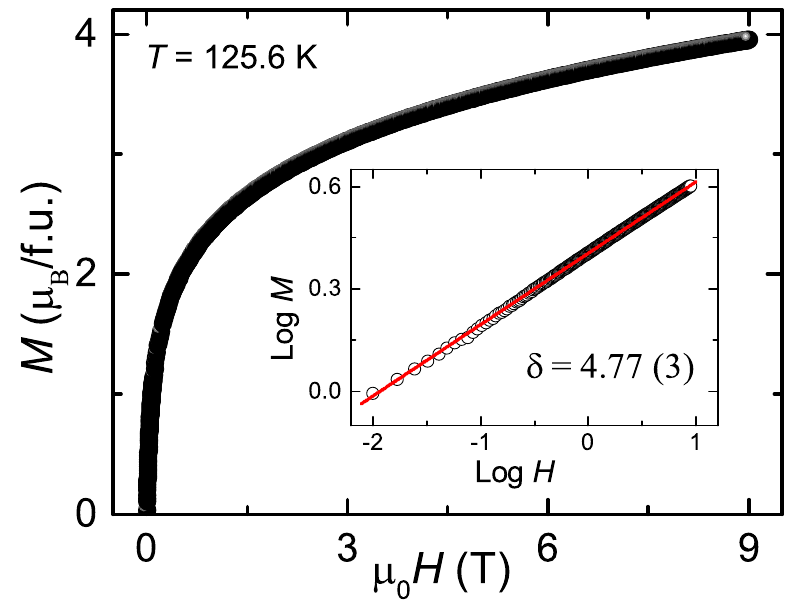}
	\caption{\label{Fig13} Magnetization isotherm ($M$ vs $H$) at $T \simeq T_{\rm C}$. Inset: log$M$ vs log$H$ plot at $T \simeq T_{\rm C}$.}
\end{figure}

On the other hand, the MAPs constructed using the critical exponents of 3D Heisenberg model resulted more linear behaviour in the high field region compared the mean-field one. Therefore, we took the theoretical values of $\beta$ and $\gamma$ for 3D Heisenberg model as the starting trial values to construct MAPs. The linear fits to the MAPs in the high field regime were extrapolated down to zero field and the values of $M_S(T)$ and $\chi_0^{-1}(T)$ were obtained from the intercepts on the $M^{1/\beta}$ and $(H/M)^{1/\gamma}$ axes, respectively. The temperature dependent $M_{\rm S}$ and $\chi_0^{-1}$ were fitted by Eqs.~\eqref{cre1} and ~\eqref{cre2}, respectively and the values of $\beta$ and $\gamma$ were estimated. These set of $\beta$ and $\gamma$ values were again used to construct a new set of MAPs. This whole process was repeated several times, until we got a set of parallel straight lines in the high field region with a set of stable $\beta$, $\gamma$, and $T_{\rm C}$ values. The final MAPs are shown in Fig.~\ref{Fig12}(b). The obtained $M_{\rm S}$ and $\chi_0^{-1}$ values are plotted as a function of temperature in Fig.~\ref{Fig12}(c). The solid lines are the fits using Eqs.~\eqref{cre1} and ~\eqref{cre2}, respectively. The fit using Eq.~\eqref{cre1} yields $\beta = 0.350(2)$ with $T_{\rm C} = 125.87(3)$~K and the fit using Eq.~\eqref{cre2} yields $\gamma = 1.320(2)$ with $T_{\rm C} = 125.83(3)$~K. These values are almost equal to the values obtained from the final MAPs ($\beta = 0.350$ and $\gamma = 1.332$). As seen from Fig.~\ref{Fig12}(b), though straight lines are obtained in the high field region but significant deviation from linearity was found in the low field region. This is likely due to the averaging over magnetic domains which are mutually misaligned. 
%Similar behavior was also found in TbCo$_{2-x}$Fe$_x$\cite{Halder174402} and Pr$_{0.5}$Sr$_{0.5}$MnO$_3$.\cite{Pramanik214426}

\begin{figure}
	\includegraphics[width=\linewidth] {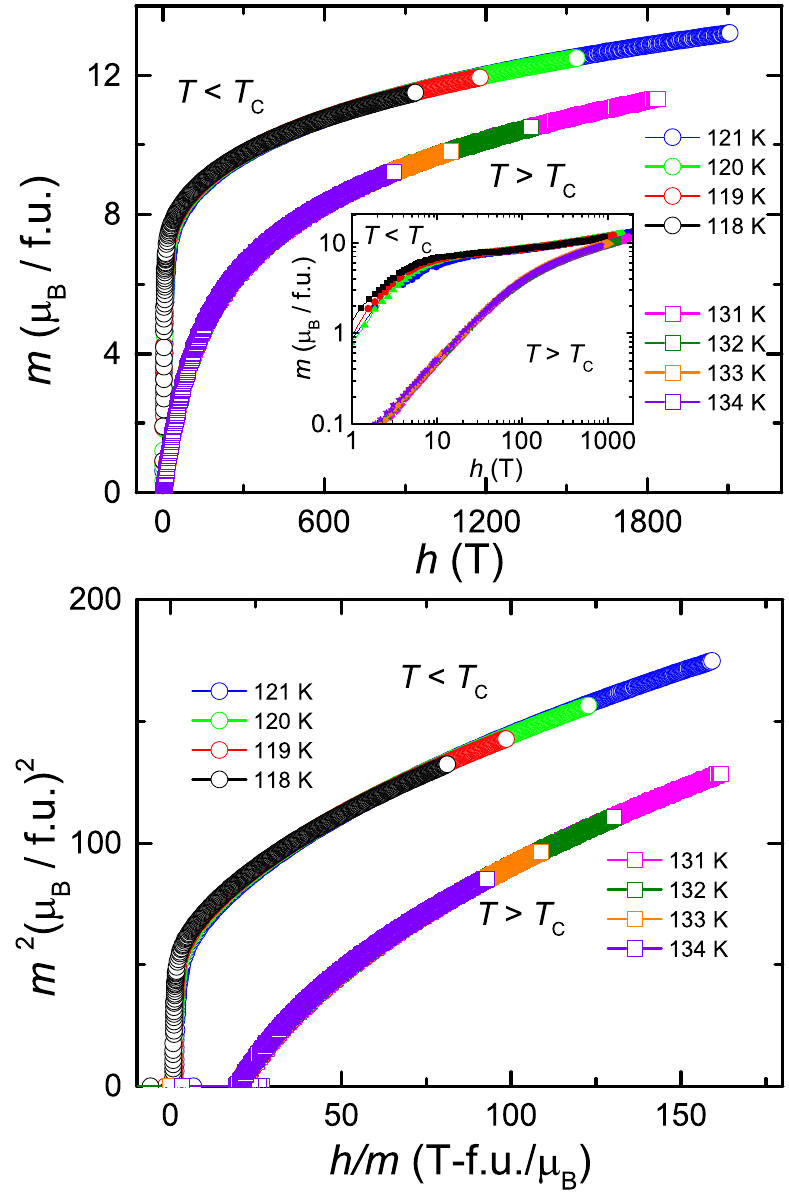}
	\caption{\label{Fig14} Upper panel: The reduced magnetization $m={\mid\varepsilon\mid}^{-\beta}M(H,\varepsilon)$ vs reduced magnetic field $h={H\mid\varepsilon\mid}^{-(\beta+\gamma)}$. The re-normalized curves in different temperatures, just above and below $T_{\rm C}$ are collapsing into two separate branches. Inset: The log-log plot of $m$ vs $h$ to magnify the data in the low field region. Lower panel: $m^2$ vs $h/m$ plots where the curves just above and below $T_{\rm C}$ are collapsing into two separate branches.}
\end{figure}

In the next step, Kouvel-Fisher (KF) method is used to determine $\beta$, $\gamma$, and $T_{\rm C}$ more accurately.\cite{KouvelA1626} The equations involved in the KF method are
\begin{equation}
M_S(T)\left [\dfrac{dM_S(T)}{dT}\right]^{-1}=(1/\beta) (T-T_{\rm C}),
\label{cre6}
\end{equation}
and
\begin{equation}
\chi_0^{-1}(T)\left [\dfrac{d\chi_0^{-1}(T)}{dT}\right]^{-1}=(1/\gamma) (T-T_{\rm C}).
\label{cre7}
\end{equation}

\begin{table*}
	\caption{Experimentally evaluated critical exponents ($\beta$, $\gamma$, and $\delta$) and $T_{\rm C}$ for LNMO obtained from MAP, KF plot, Critical Isotherm analysis, and MCE. For a comparison, the theoretically predicted values for different universality classes (mean-field, 3D Heisenberg, 3D Ising, and 3D XY) are also listed (taken from Ref.~\cite{Kaul5}).}
	\label{Tablecriticalbehabior}
	\begin{ruledtabular}
		\begin{tabular}{ p{1.2cm} p{1.4cm} p{1.4cm} p{1.4cm} p{1.4cm} p{1.0cm} p{1.0cm} p{1.0cm} p{1.0 cm} }
			Parameters  & MAP & KV plot & Critical Isotherm Analysis & MCE & Mean Field Model & 3D Heisenberg Model & 3D XY Model & 3D Ising Model\\
			\hline
			$\beta$   &0.350(2)&0.352(3)&--&--&0.5&0.365&0.345&0.325\\
			$\gamma$ &1.320(2)& 1.314(2) &--&--&1&1.386&1.316&1.241\\
			$\delta$ &4.771(2)& 4.733(4)& 4.77(3)&4.72(9)&3&4.80&4.80&4.82\\
			$n$    &0.61(2)&0.61(3) & --&0.61(2)&--&--&--&--\\
			$T_{\rm C}~(\rm K)$ & 125.83(3) &125.84(8)&125.6 &--&--&--&--&--\\
		\end{tabular}
	\end{ruledtabular}
\end{table*}

From these two equations, it is apparent that $M_{\rm S}(T) [dM_{\rm S}(T)/dT]^{-1}$ vs $T$ and $\chi_0^{-1}(T) [d\chi_0^{-1}(T)/dT]^{-1}$ vs $T$ plots should yield straight lines with slopes $1/\beta$ and $1/\gamma$, respectively. The beauty of this method is that, without any previous knowledge about $T_{\rm C}$, one can estimate it from the intercept of the straight line fits on the temperature axis. The KF plots for LNMO are presented in Fig.~\ref{Fig12}(d). The critical exponent values evaluated from the KF plots are $\beta = 0.352(3)$ with $T_{\rm C} = 125.83(6)$~K and $\gamma = 1.314(2)$ with $T_{\rm C} = 125.84(8)$~K. These values of critical exponents match nicely with the ones obtained from the MAPs, indicating the reliability of the values of critical exponents.

Figure~\ref{Fig13} presents the critical isotherm at $T \simeq T_{\rm C} = 125.6$~K. According to Eq.~$~\eqref{cre3}$, log$M$ vs log$H$ plot at the critical temperature should produce a straight line with slope $1/\delta$. We have plotted log$M$ vs log$H$ in the inset of Fig.~\ref{Fig13} and a straight line fit results $\delta = 4.77(3)$. One can also estimate $\delta$ by using the Widom relation\cite{Widom3898}
\begin{equation}
\delta = 1 + \dfrac{\gamma}{\beta}.
\label{cre8}
\end{equation}
By using the $\beta$ and $\gamma$ values from the KF method and MAPs, the corresponding $\delta$ value is estimated to be $\delta = 4.733(4)$ and $\delta = 4.771(2)$, respectively. It is found that the $\delta$ value obtained from the Widom relation and the critical isotherm analysis are very close to each other, reflecting the self-consistency of our $\beta$ and $\gamma$ estimations.

According to scaling hypothesis,\cite{Kaul5} the obtained critical exponents should follow the universal scaling function
\begin{equation}
M(H,\varepsilon)=\varepsilon^\beta f_\pm \left (\dfrac{H}{{\mid\varepsilon\mid}^{\beta+\gamma}}\right),
\label{cre4}
\end{equation}
where $f_+$ and $f_-$ are the regular functions for $T<T_{\rm C}$ and $T>T_{\rm C}$, respectively. Equation~\eqref{cre4} can further be simplified in terms of reduced magnetization ($m$) and reduced field ($h$) as 
\begin{equation}
m = f_{\pm} (h),
\label{cre9}
\end{equation} 
where $m={\mid\varepsilon\mid}^{-\beta}M(H,\varepsilon)$ and $h=H{\mid\varepsilon\mid}^{-(\beta+\gamma)}$. Equations~\eqref{cre4} and ~\eqref{cre9} suggest that for appropriate choice of critical exponents, all the curves in the $m$ vs $h$ and $m^2$ vs $h/m$ plots should fall into two separate branches: $f_+$ for $T>T_{\rm C}$ and $f_-$ for $T<T_{\rm C}$. This behaviour can be clearly visualized in Fig.~\ref{Fig14}. In the inset of the upper panel of Fig.~\ref{Fig14} we have plotted $m$ vs $h$ in log-log scale which magnify the data in the low field regime and confirms no dispersion among the curves in two branches. This further confirms the reliability of the estimated critical exponent values.

For a comparison, all the values of critical exponents evaluated by the above methods along with the theoretical values corresponding to various universality classes (mean-field, 3D Heisenberg, 3D Ising, and 3D XY) are listed in Table~\ref{Tablecriticalbehabior}. One can see that our estimated critical exponents ($\beta \sim 0.35$, $\gamma \sim 1.32$, and $\delta \sim 4.772$) are very close to the 3D XY-Model. This implies that LNMO belongs to the 3D-XY universality class. This is quite consistent with the magnetic structure deduced from the NPD data where the spin alignments are restricted only to the $ab$-plane. There are certain compounds known to show the evidence of magnetic LRO belonging to the 3D-XY universality class \textit{e.g.} CuGeO$_3$\cite {Lumsden4919}, Gd$_2$IFe$_2$, Gd$_2$ICo$_2$, and Gd$_2$BrFe$_2$\cite{Reisser265}.

\subsection{Magnetocaloric Effect}
\begin{figure*}
	\includegraphics[width=\linewidth]{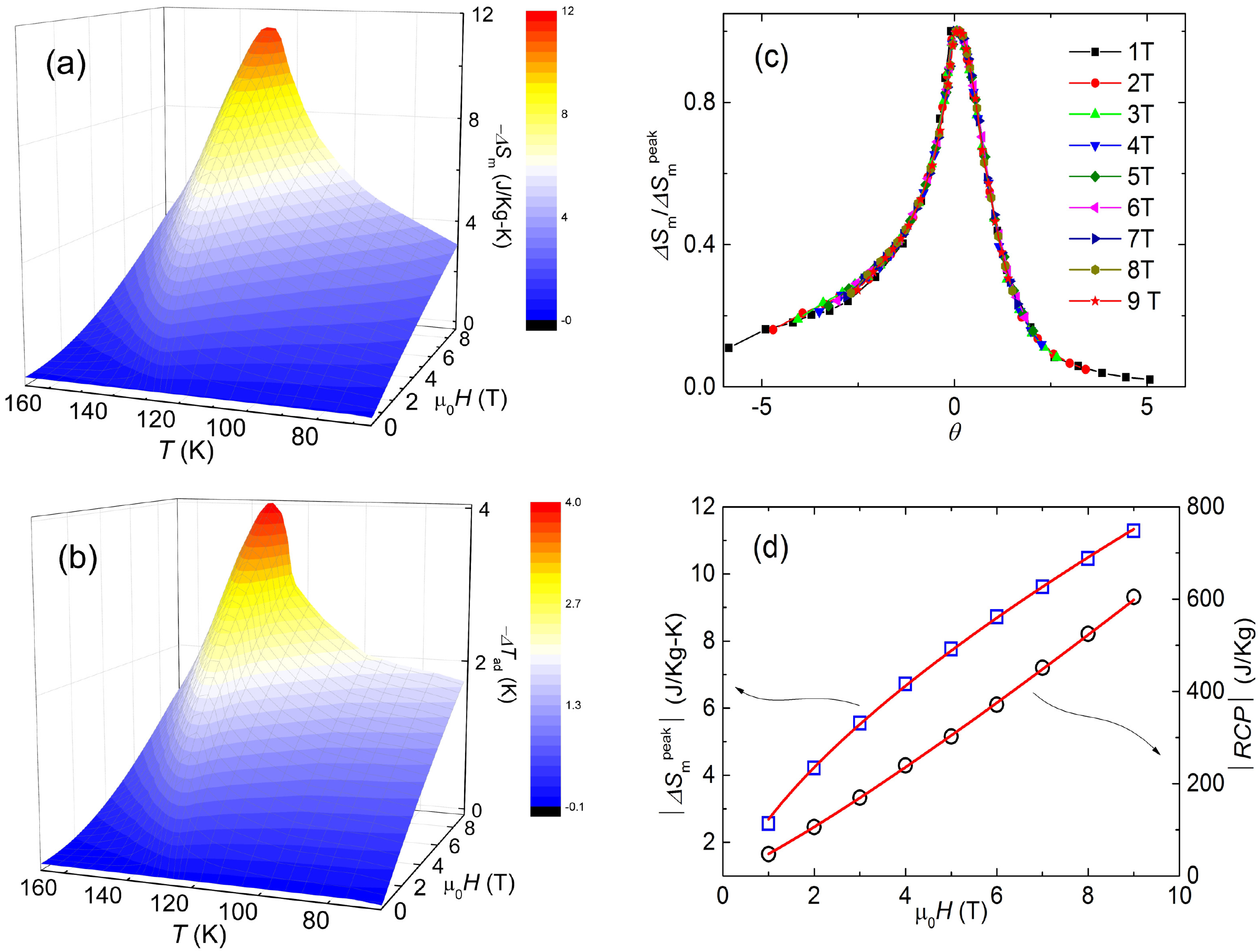}
	\caption{\label{Fig15} (a) The isothermal change in magnetic entropy ($\Delta S_{\rm m}$) as function of temperature ($T$) and field ($H$). (b) The adiabatic change in temperature ($\Delta T_{\rm ad}$) as a function of temperature ($T$) and field ($H$). (c) Universal curve plot of normalized entropy change as a function of $\theta$ with the application of 1~T to 9~T magnetic field change. (d) Absolute value of entropy change at the peak position $\Delta S_{\rm m}^{\rm peak}$ and relative cooling power ($RCP = \Delta S_{\rm m}^{\rm peak} \times \delta T_{\rm FWHM}$) as a function of magnetic field in the left and right $y$-axes, respectively. Solid lines are the fits as described in the text.}
\end{figure*}
The MCE is defined as the reversible change in temperature of a magnetic material while magnetic field is applied or removed in an adiabatic condition. Generally, isothermal entropy change ($\Delta S_{\rm m}$) and adiabatic temperature change ($\Delta T_{\rm ad}$) with respect to the change in magnetic field quantify the MCE of a system. From the Maxwell's thermodynamic relation, $(\partial S/\partial H)_T = (\partial M/\partial T)_H$, $\Delta S_{\rm m}$ can be estimated using the magnetization isotherm ($M$ vs $H$) data as
\begin{equation}
\Delta S_{\rm m} (H, T) =\int_{H_{\rm i}}^{H_{\rm f}}\dfrac{dM}{dT}dH.
\end{equation}
Figure~\ref{Fig15}(a) presents the 3D plot of $\Delta S_{\rm m}$ with the change in field ($H$) and temperature ($T$). $\Delta S_{\rm m}$ changes gradually with $H$ and shows a maximum entropy change across the $T_{\rm C}$, with a highest value of $\Delta S_{\rm m}\simeq 11.2$~J/kg-K for 9~T field change.
%Similar field dependence behavior is typically observed for materials with second order phase transition.\cite{Law2680,Singh6981}
Furthermore, $\Delta T_{\rm ad}$ is calculated from the zero field heat capacity and magnetization isotherms using the relation
\begin{equation}
\Delta T_{\rm ad}=-\int_{H_{\rm i}}^{H_{\rm f}}\dfrac{T}{C_{\rm p}}\dfrac{dM}{dT}dH.
\end{equation}
The 3D plot of $\Delta T_{\rm ad}$ as a function of $H$ and $T$ is shown in Fig.~\ref{Fig15}(b). $\Delta T_{\rm ad}$ shows a gradual temperature evolution with respect to $H$ showing a maximum value $\sim 4$~K for 9~T field change near $T_{\rm C}$. The shape of both $\Delta S_{\rm m}(T)$ and $\Delta T_{\rm ad}(T)$ peaks appear to be asymmetric and expanded over a wide temperature range around $T_{\rm C}$.

Another important parameter which is very useful to evaluate the performance of a magneto-caloric material is relative cooling power ($RCP$). It defines the amount of heat transfer between it's hot and cold reservoirs which can be written mathematically as
\begin{equation}
RCP=\int_{T_{\rm cold}}^{T_{\rm hot}}\Delta S_{\rm m}(T,H)dT.
\end{equation} 
Here, $T_{\rm cold}$ and $T_{\rm hot}$ are the temperatures corresponding to the cold and hot reservoirs, respectively. Thus, $RCP$ can be evaluated approximately as
\begin{equation}
\arrowvert RCP\arrowvert_{\rm approx}=\Delta S_{\rm m}^{\rm peak} \times \delta T_{\rm FWHM},
\end{equation}
where $\Delta S_{\rm m}^{\rm peak}$ is the peak value and $\delta T_{\rm FWHM}$ is the full-width at half maximum (FWHM) of the $\Delta S_{\rm m} (T)$ curves in Fig.~\ref{Fig15}(a). The highest value $RCP$ is found to be $\sim 604$~J/Kg for 9~T field change. This value of $RCP$ is quite high and comparable to other well known magneto-caloric materials, having $T_{\rm C}$ around this temperature.

In Table~\ref{TableMCE}, we have made a comparison of the $RCP$ and $\Delta S_{\rm m}^{\rm peak}$ values of LNMO with some other magneto-caloric materials with $T_{\rm C} = 110 - 140$~K. In most of the materials with second order phase transition, the shape of the $\Delta S_{\rm m}$ curves is broad and asymmetric which is one of the main reasons for the large value of $RCP$. Though, materials with first order phase transition show giant MCE and large $\Delta S_{\rm m}$ and $\Delta T_{\rm ad}$ values but the width of these peaks are narrow, which restricts the usability of these materials for a cyclic operation. Another drawback is that the materials with first order phase transition show hysteretic behaviour which leads to energy loss.\cite{Franco305} Therefore, materials with second order phase transition are more preferred for the magnetic refrigeration purpose than the ones with first order phase transition. It has been theoretically predicted that geometrically frustrated magnets could show enhanced MCE compared to the ordinary non-frustrated magnets.\cite{Zhitomirsky104421} In particular, pyrochlore lattices are predicted to show highest MCE among the geometrically frustrated lattices. Subsequently this idea has been utilized to design several new magnetocaloric materials with strong spin fluctuations and/or frustration.\cite{Tishin2016,Sosin094413,*Das104420} Indeed, LNMO is a frustrated pyrochlore magnet and hence the asymmetric behaviour can be attributed to the effect of magnetic frustration.
Thus, the obtained large $\Delta S_{\rm m}$ and $RCP$ values make LNMO a promising compound for magnetic refrigeration purpose. 
\begin{table*}
	\caption{Comparison of experimental $\Delta S_{\rm m}^{\rm peak}$ and $RCP$ values for LNMO with some reported magneto-caloric materials having $T_{\rm C} = 110 - 140$~K for a field change of 5~T.}
	\label{TableMCE}
	\begin{ruledtabular}
		\begin{tabular}{ccccccccc}
			system &  $ T_{\rm C}$ (K)  & Nature of transition &  $ \Delta S_{\rm m}^{\rm peak} $ (J/Kg-K) & $RCP$ (J/Kg) & Ref. \\  \hline
			Tb$_2$NiMnO$_6$ & 110 & Second order & 5.2 & - &\cite{Chakraborty59}\\
			DyGa & 113 & Second order & 5.8 & 381.9 &\cite{Zheng07A917}\\
			LiNi$_{0.5}$Mn$_{1.5}$O$_4$ & 125.8 & Second order & 7.76 & 302 & This work \\
			GdCo$_{0.2}$Mn$_{1.8}$ & 140 & Second order & 4.11 & 320 & \cite{Zhang541} \\
			Tb$_5$Ge$_{2-x}$Si$_{2-x}$Mn$_{2x}$, $2x = 0.1$ & 123 & First order & 20.84 & 330.43 &\cite{Yuzuak057501}\\
			Dy(Co$_{0.98}$ Ni$_{0.02}$)$_{2}$ & 126 & First order & 11 & 304 & \cite{Balli7601} \\
			YFe$_{2}$H$_{4.2}$ & 132 & First order & 7.11 & 263 &\cite{Paul-Boncour013914}\\
		\end{tabular}
	\end{ruledtabular}
\end{table*}

Further, MCE can also be utilized to gain more insight about the nature of the magnetic phase transition and the critical exponents. The universal scaling curve construction of $\Delta S_{\rm m}$ proposed by Franco~\textit{et.~al.}\cite{Franco222512,*Franco093903} is generally used for this purpose. This method is tested on variety of materials and found to be a very efficient way of investigating the nature of the phase transition.\cite{Bonilla224424} To construct the universal scaling curves, first we normalized all the $\Delta S_{\rm m}$($T$) curves with their respective peak values (i.~e.~$\Delta S_{\rm m}$/$\Delta S_{\rm m}^{\rm peak}$) for each field change and then plotted as a function of $\theta$ in Fig.~\ref{Fig15}(c). Here, $\theta$ is the re-scaled temperature, which is given by
\begin{equation}
\theta=-(T-T_{\rm C})/(T_{\rm r1}-T_{\rm C})~{\rm for}~T\leq T_{\rm C}
\label{}
\end{equation}
and
\begin{equation}
\theta=~(T-T_{\rm C})/(T_{\rm r2}-T_{\rm C})~~{\rm for}~T>T_{\rm C}.
\label{}
\end{equation}
In the above, $T_{\rm r1}$ and $T_{\rm r2}$ are the two reference temperatures corresponding to $0.5 \times \Delta S_{\rm m}^{\rm peak}$ values and $T_{\rm C}\simeq 125.8$~K [obtained from the Kouvel-Fisher plot]. As it is seen from Fig.~\ref{Fig15}(c), all the normalized curves for different field changes collapse into a single curve similar to the other reported compounds showing second order phase transition.\cite{Bonilla224424,Singh6981}

The $\Delta S_{\rm m}^{\rm peak}$ and $RCP$ values are plotted in Fig.~\ref{Fig15}(d), as a function of field in the left and right $y$-axes, respectively. For the purpose of critical analysis, we have fitted the field dependent $\Delta S_{\rm m}^{\rm peak}$ and $RCP$ curves by the following power laws\cite{Franco222512,*Franco093903}
\begin{equation}
\lvert\Delta S_{\rm m}^{\rm peak}\arrowvert \propto H^n
\label{25}
\end{equation}
and
\begin{equation}
\arrowvert RCP \arrowvert \propto H^{(1+1/\delta)}.
\label{26}
\end{equation}
The fit of Eq.~\eqref{25} to the $\Delta S_{\rm m}^{\rm peak}$ vs $T$ data yields $n = 0.61(2)$. The exponent $n$ is related to the critical exponents $\beta$ and $\gamma$ by the relation
\begin{equation}
n=1+\left (\dfrac{\beta-1}{\beta+\gamma}\right).
\label{Eq.31}
\end{equation} 
Using $\beta$ and $\gamma$ values obtained from the modified Arrot plot and Kouvel-Fisher plots, the value of $n$ is estimated to be 0.61(2) and 0.61(3), respectively. These values of $n$ obtained from various analysis methods are consistent with each other and confirms the reliability of our critical analysis technique. Similarly, the fit of $RCP$ vs $H$ data using Eq.~\eqref{26} yields $\delta = 4.72(9)$, which is consistent with the values obtained from modified Arrrott plot, Kovel-Fisher plot, and critical isotherm analysis techniques.

\begin{figure}
	\includegraphics[width=\linewidth] {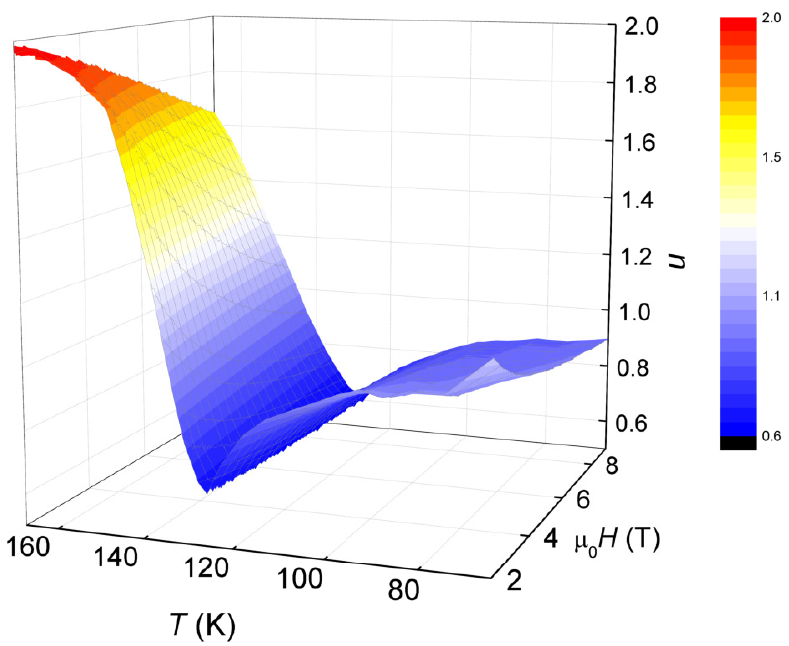}
	\caption{\label{Fig16} Field and temperature dependence of the exponent $n$ using Eq.~\eqref{Eq.33}.}
\end{figure}
A more quantitative analysis of the phase transition can be done by using the method proposed recently by Law \textit{et.~al.}\cite{Law2680}. The exponent $n$ in Eq.~\eqref{25} is normally field and temperature dependent and can also be calculated locally as
\begin{equation}
n(T,H) = \dfrac{{\rm d~ln}\arrowvert\Delta S_{\rm m}\arrowvert}{{\rm d~ln}H}.
\label{Eq.33}
\end{equation}
We used Eq.~\eqref{Eq.33} to quantify the local $H$ and $T$ variation of exponent $n$ and plotted as a 3D plot in Fig.~\ref{Fig16}. Since in the low field range, the system has multi-domain states, we have shown only the values corresponding to high fields ($H>2$~T) in Fig.~\ref{Fig16}. At low temperatures ($T < T_{\rm C}$), the $n$ vs $T$ curve is found to be almost constant and $n$ approaches a value $\sim 0.94$ which is close to 1. As the temperature is increased, $n$ decreases smoothly and passes through a minima at $T \simeq T_{\rm C}$ with $n = 0.61(1)$. At high temperatures, it increases almost linearly and reaches upto a value $n = 1.98$ which is close to 2. At $T = T_{\rm C}$, the value of $n$ depends on the value of other critical exponents of the material, as they are connected by Eq.~\eqref{Eq.31}. The overall behaviour of $n$ vs $T$ curve (with exponent value, $n<2$) is found to be similar to other compounds showing second order phase transition.\cite{Law2680}
% \begin{figure}
%	\includegraphics[width=\linewidth] {Fig13}
%	\caption{\label{Fig13} Relative cooling power ($RCP = \Delta S_{\rm m}^{\rm peak} \times \delta T_{\rm FWHM}$) and absolute value of entropy change at the peak position $\Delta S_{\rm m}^{\rm peak}$ as a function of magnetic field in the left and right $y$-axes, respectively. Solid lines are the fits as described in the text.}
%\end{figure}

\subsection{AC Susceptibility}
\begin{figure}
	\includegraphics[width = \linewidth] {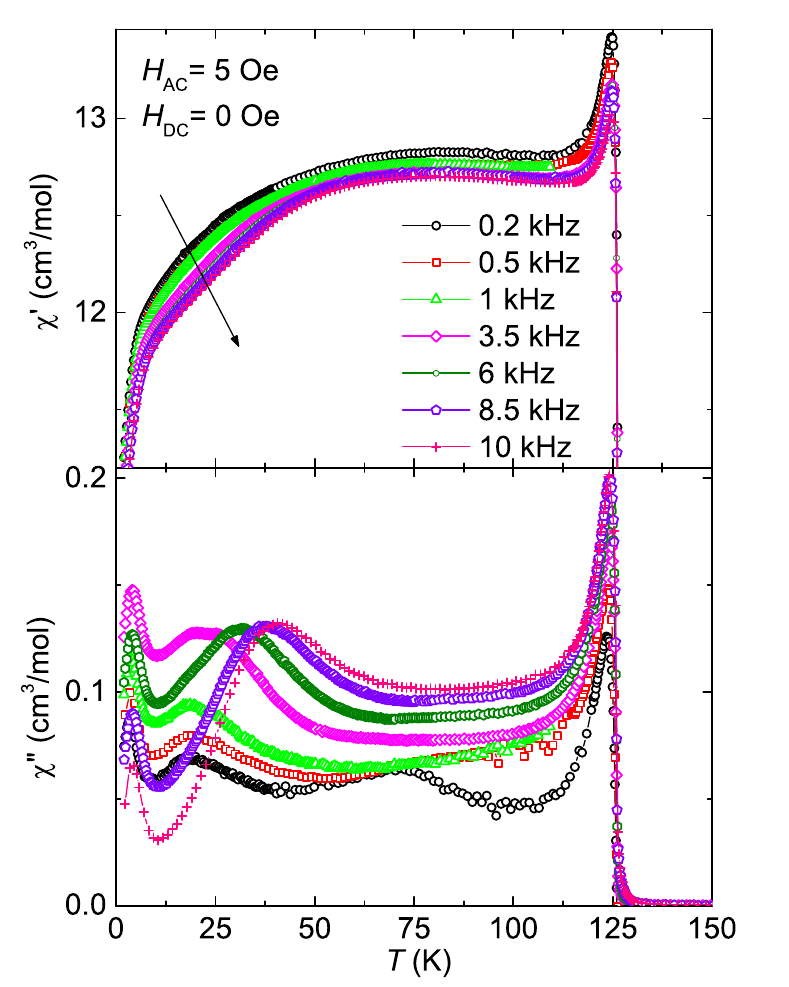}
	\caption{\label{Fig17} Upper panel: Real part of the AC susceptibility $\chi'$ vs temperature at different frequencies. The arrow points to the change in $\chi'$ with frequency. Lower panel: Imaginary part of the AC susceptibility $\chi''$ vs temperature at different frequencies.}
\end{figure}
\begin{figure}
	\includegraphics[width= \linewidth] {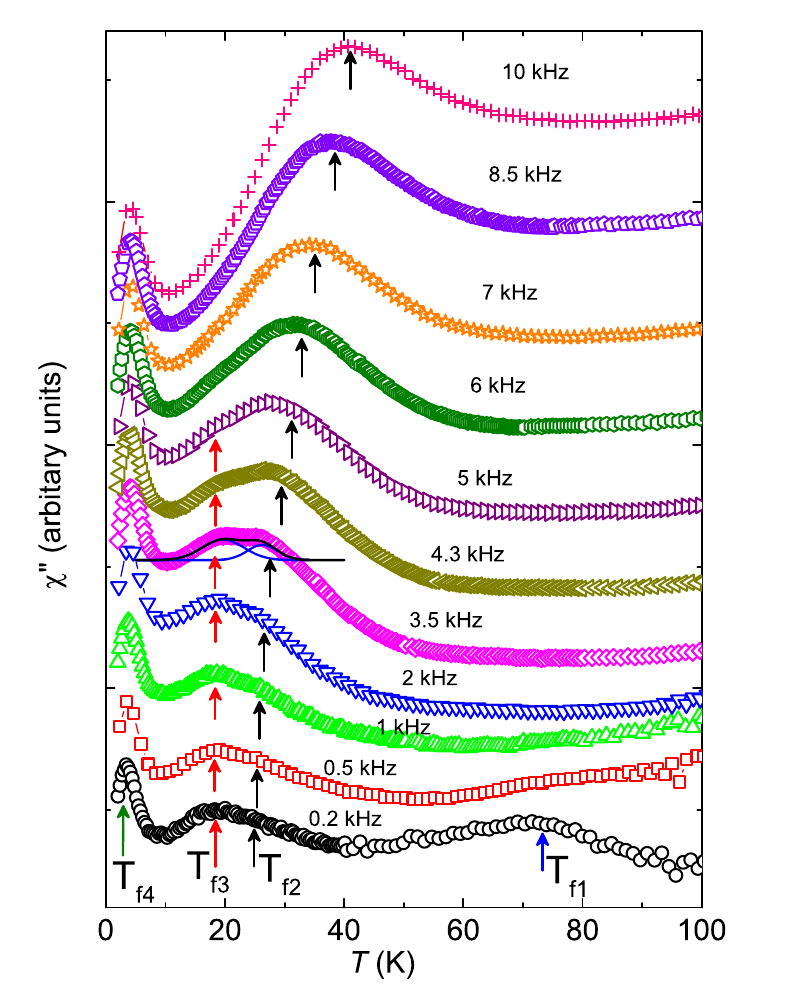}
	\caption{\label{Fig18} Enlarged view of $\chi^{''}(T)$ at low temperatures and at different frequencies. The curves are vertically translated to get a clear view of the peak positions. The vertical arrows guide the peak positions with frequency.}
\end{figure}
Figure~\ref{Fig17} presents the temperature dependent AC susceptibility measured in different frequencies and at a fixed AC field of $H_{\rm AC}\simeq$ 5 Oe. As shown in the upper panel of Fig.~\ref{Fig17}, the real part of the AC susceptibility ($\chi^{\prime}$) shows a sharp peak at $T_{\rm C} \simeq 125$~K and below 50~K it decreases with temperature, akin to the behaviour observed in the ZFC DC $\chi(T)$ (inset of Fig.~\ref{Fig5}). The peak at 125~K is almost frequency independent indicating the onset of long-range ferrimagnetic order. Another broad feature appears at around 25~K in low frequencies and the absolute value of $\chi^{\prime}$ at this temperature is found to be reduced with increasing frequency (indicated by a downward arrow). The imaginary part of the AC susceptibility ($\chi^{''}$) is shown in lower panel of Fig.~\ref{Fig17} as a function of temperature. Similar to the $\chi^{\prime}(T)$ data, $\chi^{''}(T)$ also shows a sharp peak at $T_{\rm C}$ which is frequency independent. At low temperatures, $\chi^{''}(T)$ shows various anomalies and some of them are frequency dependent.

To clearly visualize those low temperature features, $\chi^{''}(T)$ data at different frequencies are vertically translated and magnified in Fig.~\ref{Fig18}. At the lowest measured frequency of 0.2~kHz, $\chi^{''}(T)$ below $T_{\rm C}$ exhibits two broad humps at $\sim 72$~K and $\sim 20$~K and a narrow peak at $\sim 3.5$~K, as indicated by the upward arrows. The broad hump at $T_{\rm f1} \sim 72$~K disappears as the frequency is increased beyond 0.5~kHz. With a very careful observation, we found that the broad hump near 20~K consists of two shoulder like features, one at $T_{\rm f2}\simeq 18$~K and another at $T_{\rm f3}\simeq 20$~K. As the frequency is increased, the shoulder at $T_{\rm f2}$ becomes more pronounced and shifts towards high temperatures, reflecting a glassy behaviour. On the other hand, the shoulder at $T_{\rm f3}$ remains frequency independent and gets diminished at high frequencies. Similarly, the peak at $T_{\rm f4} \sim 3.5$~K also remains unaltered with frequency indicating that it could be an AFM transition. 
%A glass transition in the magnetic ordered state is the reminiscent of a reentrant spin-glass (RSG) behaviour.\cite{Manna224420,Hanasaki086401,Kumar011037,Jonason6507}The development of a RSG state is often observed as a consequence of frustration arising from the competing exchange interactions within a LRO state.
The multiple magnetic transitions in the ordered state can be attributed to the effect of magnetic frustration.\cite{Dho027202}
%similar to that observed in the Cr doped perovskite manganite.\cite{Dho027202}
The absence of these features in $\chi^{\prime}$ can be believed as a consequence of large value of $\chi^{\prime}$ associated with ferrimagnetism.\cite{Mahendiran104402}

Since all the peaks are frequency independent except one, we focused our analysis only on the frequency dependent peak at $T_{\rm f2}$. The peak positions of $T_{\rm f2}$ and $T_{\rm f3}$ are extracted by a double Gaussian fit (shown in Fig.~\ref{Fig18} for 3.5 kHz). The relative shift of freezing temperature ($T_{\rm f2}$) with frequency ($\nu$) can be quantified by calculating the Mydosh parameter $\delta T_f$, given by\cite{Mydosh2014}
\begin{equation}
\delta T_f = \left (\dfrac{\Delta T_{\rm f}}{T_{\rm f} \Delta (\rm log_{10}\nu)}\right),
\end{equation}
where $\Delta T_{\rm f} = (T_{\rm f})_{\nu_1} - (T_{\rm f})_{\nu_2}\simeq16.65$~K and $\Delta \rm log_{10}(\nu) = \rm log_{10}(\nu_1) - \rm log_{10}(\nu_2)$. Here, $\nu_1$ and $\nu_2$ are taken as the lowest (200~Hz) and highest (10~kHz) measured frequencies, respectively. For LNMO, the Mydosh parameter is estimated to be $\delta T_f \simeq 0.4$, which is found to be two orders of magnitude higher than the value for canonical SG systems [\textit{e.g.} AuMn, ($\delta T_{\rm f}\simeq 0.0045$)]\cite{Mulder515} and one order magnitude higher than the cluster SG systems [\textit{e.g.} Cr$_{0.5}$Fe$_{0.5}$Ga, ($\delta T_{\rm f}\simeq 0.017$)].\cite{Bag144436} However, it lies in the range of a superparamagnet [\textit{e.g.} $\alpha-$(Ho$_2$O$_3$B$_2$O$_3$), ($\delta T_{\rm f}\simeq$ 0.28)].\cite{Mydosh2014}
\begin{figure}
\includegraphics[width= \linewidth] {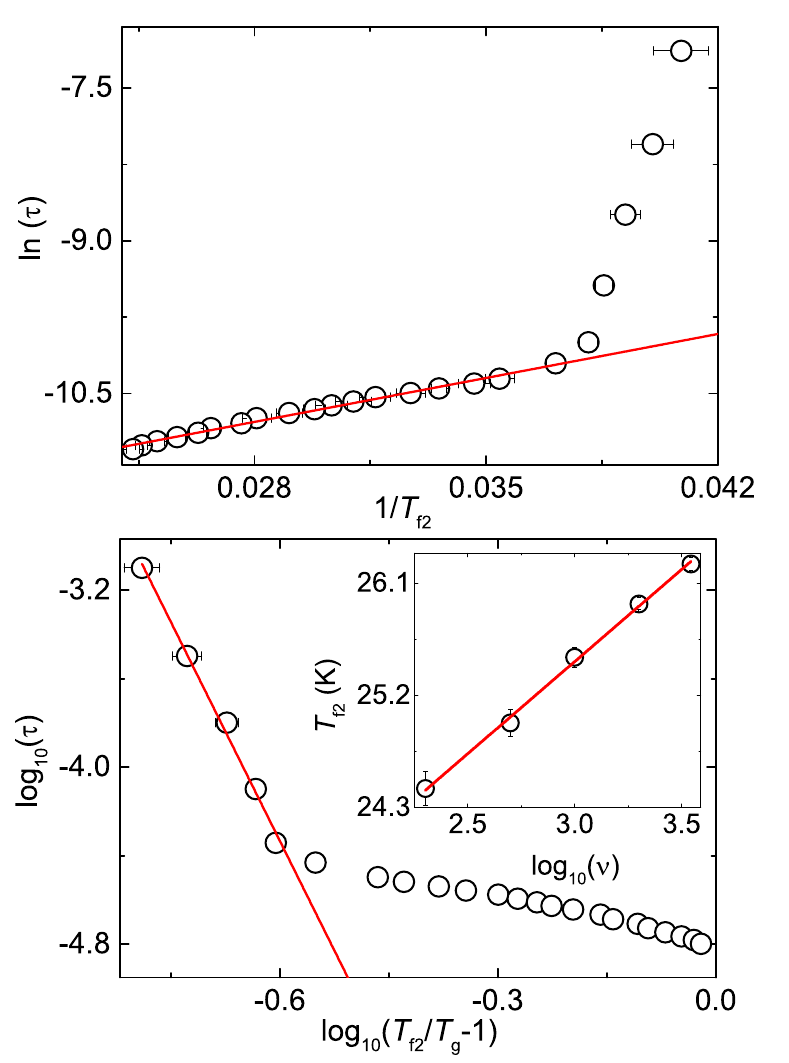}
\caption{\label{Fig19} Upper panel: ln$(\tau)$ vs 1/$T_{\rm f2}$ plot. The solid line is the fit using Eq.~\eqref{arrhenius_law} in the high temperature region. Lower panel: $\rm log_{10}(\tau)$ vs $\rm log_{10} (T_{\rm f2}/T_{\rm g}-1)$ plot. The solid line is the fit using Eq.~\eqref{pwlaw_2}. Lower inset: $T_{\rm f2}$ vs $\rm log_{10}\nu$ plot in the low temperature region. The solid line is the fit as described in the text.}
\end{figure}

To understand the frequency dependence of $T_{\rm f2}$ and the correlation among the magnetic entities we analyzed the data using various theoretical models. At first, we tried to fit the data using Arrhenius law\cite{Binder801}
\begin{equation}
\tau = \tau_0~{\rm exp} \left (\dfrac {E_{\rm a}}{ k_{\rm B} T_{\rm f}}\right),
\label{arrhenius_law}
\end{equation}
where $\tau$ is the time scale of dynamical fluctuation, $\tau_0$ is the relaxation time for the single spin-flip, $E_{\rm a}/k_{\rm B}$ is the activation energy required to overcome the energy barrier by which the meta-stable states are separated, and $\tau_0$ is the time taken to overcome the energy barrier. In the upper panel of Fig.~\ref{Fig19}, we have plotted ln$\tau$ vs $1/T_{\rm f2}$. Clearly, it shows a change in slope at around 27~K. In the high temperature ($T>27$~K) region, the data could be fitted well by Eq.~\eqref{arrhenius_law} resulting $E_{\rm a}/k_{\rm B}\simeq 61.7$~K and $\tau_0 \simeq 3.7 \times 10^{-6}$~s. Such an Arrhenius behaviour is often considered to be the characteristic feature of superparamagnetism.\cite{Suzuki104418} The value of $\tau_0$ is also seems to be within range expected for superparamagnets ($10^{-6}$ to $10^{-9}$ s)\cite{Kumar144409}.

In the low temperature ($T<27$~K) region, the data show a significant deviation from the Arrhenius law as evident in the upper panel of Fig.~\ref{Fig19}. This indicates that there are two different relaxation mechanisms involved. We fitted the low-$T$ data by the dynamical scaling law or power law, predicted for SG systems\cite{Souletie516}
\begin{equation}
\tau = \tau'~{\rm exp} \left(\dfrac{T_{\rm f2}}{ T_{\rm g}}-1\right)^{-z\nu'}.
\label{pwlaw_1}
\end{equation}
Here, $\tau'$ has the same physical meaning as $\tau_0$, $T_{\rm g}$ is the freezing temperature as $\nu$ approaches zero, and $\tau'$ = $\xi^{\rm z}$ with $z$ being the dynamical critical exponent. The correlation length has the form $\xi = (T_{\rm f2}/T_{\rm g}-1)^{-\nu'}$ with critical exponent $\nu'$. Equation~\eqref{pwlaw_1} can be rewritten in a simplified form as
\begin{equation}
{\rm log_{10}}\tau = {\rm log_{10}}\tau'-z\nu'{\rm log_{10}}\left(\dfrac{T_{\rm f2}}{ T_{\rm g}}-1\right).
\label{pwlaw_2}
\end{equation}
As shown in the lower panel of Fig.~\ref{Fig19}, Eq.~\eqref{pwlaw_2} fits well to the low-$T$ data giving $\tau' \simeq 5.08\times10^{-9}$~s and $z \nu'\simeq(6.6\pm0.2)$. In the fitting process we have fixed $T_{\rm g}\simeq(21.03\pm0.01)$~K, obtained from the $y$-intercept of the linear fit of $T_{\rm f2}$ vs log$_{10}\nu$ plot (see the inset of lower panel of Fig.~\ref{Fig19}). The larger value of $\tau'$ clearly indicates that the spin dynamics is slower than conventional SG systems ($\sim 10^{-13}$ s).\cite{Mydosh2014} Such a high value of $\tau'$ is previously reported for various reentrant-SG systems.\cite{Kumar144409,Viswanathan012410,*Hanasaki086401} Further, the value of z$\nu'$ also falls in the range expected for typical SG systems ($z\nu'\simeq 4 \textendash 12$) and comparable to the values reported for various reentrant-SG systems.\cite{Kumar144409,Viswanathan012410,*Hanasaki086401} 

Thus, the frequency dependence of $T_{\rm f2}$ follows an Arrhenius behaviour at higher temperatures  which could suggest that the system is superparamagnet far above the critical region ($T \gg T_{\rm g}$). As we move closer to $T_{\rm g}$, Arrhenius behaviour breaks down and power law behaviour takes over which evidences the existence of SG transition at $T_{\rm g} \simeq 21.03$~K, similar to the dilute magnet LiHo$_{x}$Y$_{1-x}$F$_{4}$ ($x=0.045$).\cite{Quillium187204} In order to resolve this ambiguity, memory effect measurements are discussed later.
%Furthermore, nano sized particles in ferromagnetic domains can also exhibit glassy magnetic features.\cite{Mahendiran104402,Wu174408} At this point it is hard to tell that the low temperature ground state is a SG or a superparamagnetic state. To clarify this issue, other experimental tests such as aging and magnetic memory effect measurements are required.
 
%There is negative behavior of $\chi''(T)$ can be seen with decreasing temperature for 200~Hz and 500~Hz at $T<130$~K shown in the lower panel of Fig.~\ref{Fig17}. similar behavior is also observed for few SG compounds SrSn$_{2}$Fe$_{4}$O$_{11}$ \cite{Shlyk054426} and BiFeO$_{3}$\cite{Singh144403}. The reason behind this phenomena is not fully understood and requires more detailed investigation.
\subsection{Nonequilibrium Dynamics}
\subsubsection{Magnetic Relaxation}
\begin{figure}
	\includegraphics[width = \linewidth] {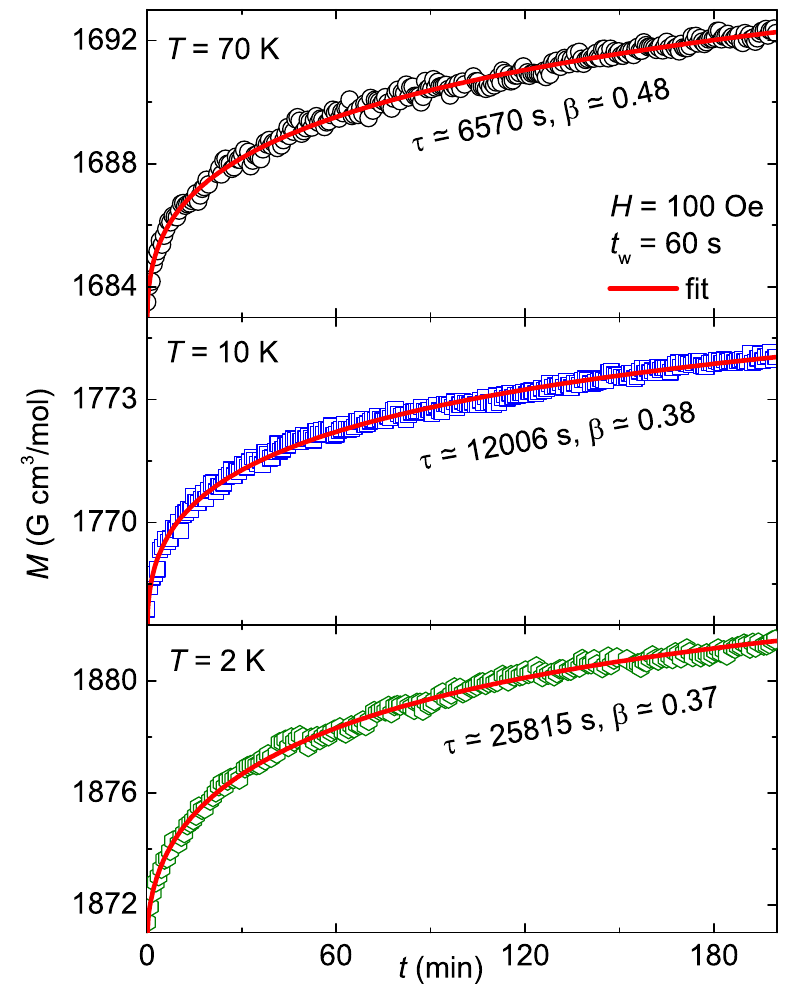}
	\caption{\label{Fig20} Time dependence of ZFC magnetization at different temperatures ($T = 2, 10, \rm {and}~70$~K) in $H = 200$~Oe and after a waiting time of 60~s. The solid lines are the fits using Eq.~\eqref{strchd eq}.}
\end{figure}
To explore the low temperature spin dynamics in LNMO, we have performed the magnetic relaxation time measurement at different temperatures (2~K, 10~K, and 70~K), below $T_{\rm b}$. The sample was cooled in ZFC condition from high temperature paramagnetic state ($T\geq 200$~K) to the desired temperature. After waiting for $t_{\rm w} = 60$~s at the desired temperature, a small magnetic field of $H \simeq 100$~Oe is applied. Thereafter, the growth of the magnetization with time [$M(t)$] was recorded. The resulting $M(t)$ curves are plotted in Fig.~\ref{Fig20}.
The time evolution of magnetization curves are found to follow a stretched exponential behaviour specified by the Kohlrausch-Williams-Watts relation\cite{Kroder174410,Alvarez7306}
\begin{equation}
M(t) = M_{\rm 0} - M_{\rm g}~{\rm exp}[-(t/\tau)^{\beta}].
\label{strchd eq}
\end{equation}
Here, $M_{\rm 0}$ is the intrinsic magnetization at $t = 0$, $M_{\rm g}$ is associated with the glassy component of the magnetization, $\tau$ is the characteristic relaxation time, and $\beta$ is the stretching exponent. Typically, the value of $\beta$ varies between 0 and 1 which decides the dynamics of a spin system. $\beta$ is also a function of temperature and is strongly dependent on the nature of energy barrier involved in the relaxation process. Following Eq.~\eqref{strchd eq}, when $\beta = 0$, $M(t) =$ constant; means no relaxation. Similarly, $\beta = 1$ implies relaxation of a spin system with a single time constant due to the presence of uniform energy barrier. On the other hand, $\beta<$~1 implies the presence of the distribution of non-uniform energy barriers, typically observed for SG and superparamagnetic systems.\cite{Bag144436,Kroder174410,De033919,Tsoi014445}

The $M(t)$ curves for $T= 70$~K, 10~K, and 2~K are well fitted by Eq.~\eqref{strchd eq}, yielding $\beta \simeq 0.48$, 0.38, and 0.37, respectively. These values are less than 1 suggesting the evolution of magnetization through a number of intermediate metastable states. The value of $\tau$ increases with decreasing temperature as expected for glassy systems.\cite{Bag144436,Li7434} The relaxation behaviour observed at $T = 10$~K and $T = 2$~K is quite natural because both of these temperatures are well below $T_{\rm f2}$. However, the slow relaxation behaviour observed at high temperatures ($T = 70$~K~$>$~$T_{\rm f2}$) is unusual and suggests that the persistence of SG/superparamagnetism beyond $T_{\rm f2}$ in the ferrimagnetically ordered state.
%This seems to be consistent with the multiple glassy transitions observed in the AC $\chi(T)$ data.

\subsubsection{Magnetic Memory Effect}

\begin{figure}
	\includegraphics[width = \linewidth] {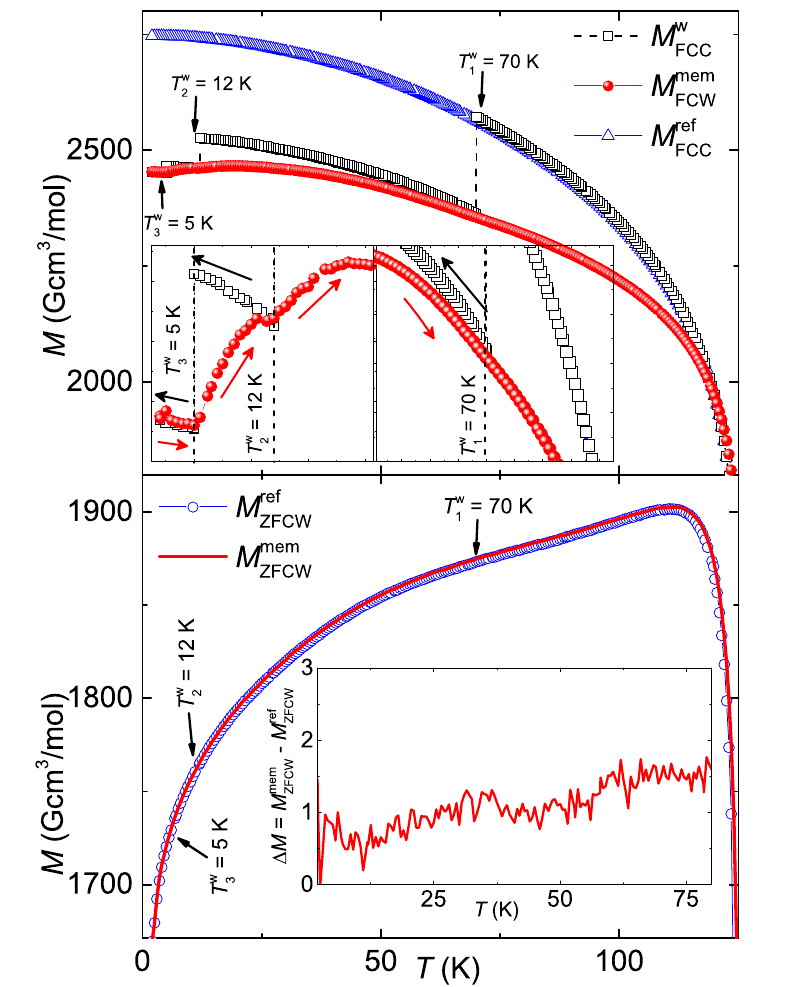}
	\caption{\label{Fig21} Memory effect measured as a function of temperature in FC (upper panel) and ZFC (lower panel) protocols in $H = 100$~Oe, as described in the text. Inset of the upper panel magnifies the features at the interruption points: $T=5$~K, 12~K, and 70~K. The difference in magnetization, $\Delta M (= M^{\rm mem}_{\rm ZFCW}- M^{\rm ref}_{\rm ZFCW})$ vs $T$ is plotted in the inset of the lower panel.}
\end{figure}

Since the bifurcation of the DC ZFC and FC $\chi(T)$ data occurs for both SG and superparamagnetic systems, the history-dependent measurements are required in order to delineate the microscopic character of the spin system. Therefore, in the following, we have described the magnetic memory effect measurements in both FC and ZFC protocols.

In the FC protocol, the sample was cooled in a small magnetic field of $H = 100$~Oe from high temperature paramagnetic state ($T\geq 200$~K) to the lowest measured temperature $T = 2$~K at a constant rate of 0.5~K/min with intermediate stops at three different temperatures ($T_{1}^{\rm w} = 70$~K, $T_{2}^{\rm w} = 12$~K, and $T_{3}^{\rm w} = 5$~K). At each stop, the magnetic field was switched off and a waiting time of $t_{\rm w} = 3$~hours was given for the magnetization to relax. After each $t_{\rm w}$, the same magnetic field was applied and field cooled cooling (FCC) process was resumed. In this way, the recorded magnetization ($M^{\rm w}_{\rm FCC}$) is plotted in the upper panel of Fig.~\ref{Fig21} which shows step like features at each stopping temperature. Once it reached 2~K, the magnetization was recorded by heating the sample at the same rate (0.5~K/min) back to 200~K in the same magnetic field without any intermediate stop. The resulting field cooled warming magnetization data are referred as $M^{\rm mem}_{\rm FCW}$ in the upper panel of Fig.~\ref{Fig21}. As depicted in the inset of the upper panel of Fig.~\ref{Fig21}, $M^{\rm mem}_{\rm FCW}$ also exhibits change of slope at each stopping temperature ($T_{2}^{\rm w} = 12$~K and $T_{3}^{\rm w} = 5$~K) as in $M^{\rm w}_{\rm FCC}$. These characteristic features clearly imply that the system tries to remember the thermal history of magnetization during cooling, thus, showing the magnetic memory. A weak change of slope was also seen at the stopping temperature, $T_{1}^{\rm w} = 70$~K which is well above $T_{\rm f2}$. Similar behaviour is also reported for reentrant SG compound Lu$_2$MnNiO$_{6}$ and it is mentioned that 
the unusual behaviour where memory exists above $T_{\rm f}$ and below $T_{\rm C}$ can be attributed to the effect of magnetic frustration due to competing FM and AFM interactions.\cite{Manna224420}
%The observation of slow relaxation and magnetic memory effect at high temperatures ($T > T_{\rm g}$) is a possible indication of the presence of glassy state above $T_{\rm f2}$. Similar behavior is also reported for other re-entrant glassy systems such as Lu$_2$MnNiO$_{6}$.\cite{Manna224420}
A FCC curve, $M^{\rm ref}_{\rm FCC}$ in the same field without any interruption is also measured as a reference.

In the ZFC protocol, the sample was cooled in zero magnetic field form $T\geq200$~K to 2~K at a constant rate of 0.5~K/min with three intermediate stops at $T_{1}^{\rm w} = 70$~K, $T_{2}^{\rm w} = 12$~K, and $T_{3}^{\rm w} = 5$~K. At each stop, the sample was allowed to relax for a waiting time of 3~hours. After reaching 2~K, the magnetization $M^{\rm mem}_{\rm ZFCW}$ was collected by warming the sample upto 200~K, after applying a small magnetic field of 100~Oe. The reference data, $M^{\rm ref}_{\rm ZFCW}$ were also taken by measuring the magnetization during warming in the same magnetic field 100~Oe, after the sample was cooled in zero magnetic field without any intermediate stops. From the data presented in the lower panel of Fig.~\ref{Fig21}, it can be seen that there is neither any clear dip nor any change in slope in the $M^{\rm mem}_{\rm ZFCW}$ data at the stopping temperatures, implying the absence of ZFC memory effect. The difference in magnetization, $\Delta M (= M^{\rm mem}_{\rm ZFCW}- M^{\rm ref}_{\rm ZFCW})$ vs $T$ is also plotted in the inset of the lower panel of Fig.~\ref{Fig21} to highlight no memory effect. In order to make sure that there is no ZFC memory, we have also studied memory effect by measuring AC susceptibility in the ZFC protocol, following the same procedure as discussed above. Similar to the DC $\chi(T)$, the real ($\chi'$) and imaginary ($\chi''$) parts of the AC susceptibility data (not shown) don't show any change of slope at the stopping temperatures. This further proves that no memory is imprinted by aging under zero field.

The above memory effect can be understood from the simple two-state model proposed by Sasaki~$et.~al.$\cite{Sasaki104405} and Tsoi~$et.~al.$\cite{Tsoi014445} for non-interacting magnetic nano particles (superparamagnets). In this model, it is assumed that a superspin associated with the dipole magnetic moment of a nano particle can occupy one of the two states with energies $-KV\pm HM_{\rm s}V$, where $K$ is the bulk anisotropy constant, $V$ is the volume of the nanoparticle, and $H$ is the applied field. Therefore, a broad distribution of particle volumes results in a broad distribution of anisotropic energy barriers. The occupation probability $p_{1}(t)$ of one of the two states, in which the superspin is antiparallel or parallel to the applied field is $p_{1}(t)= 0.5$ [i.e. $M(t)=0$] at any time $t$, if $p_{1}(t=0)= 0.5$ and $H=0$. Thus, in a ZFC process, which starts from a initial demagnetized state [$M (t=0)= 0$], $p_{1}(t)$ and hence the total magnetization is independent of the waiting time ($t_w$), whereas for a FC process which starts from a initial magnetized state, $p_{1}(t)$ and hence the total magnetization is dependent on $t_w$. In the light of this model, a superparamagnet should not show ZFC memory but it can show FC memory simply because of the blocked (frozen) superspins. On the other hand, a SG system can show both FC and ZFC memories which are well explained by Sasaki~$et.~al.$\cite{Sasaki104405} considering the random energy model\cite{Derrida2613,*Bouchaud1705} as well as the droplet theory\cite{Fisher373} proposed for SG systems. The above models have been employed to describe the experimental data of various superparamagnets and SG systems.\cite{Sasaki104405,Tsoi014445,Bandyopadhyay214410,Chen214436} Thus, the absence of ZFC memory discriminates the dynamics of LNMO from the behaviour of a SG and establishes the superparamagnetic nature at low temperatures.\cite{Tsoi014445,De033919} %Further, since the compound undergoes a ferrimagnetic transition at $T_{\rm C}$, the above observations point to the existence of superparamagnetism in the long-range ferrimagnetic ordered state.
It is quite surprising that despite having large average particle sizes ($\sim 100$~nm), the compound still behaves like a superparamagnet. It is to be noted that the unconventional superparamagnetism is also reported in several compounds in polycrystalline form.\cite{Bajpai8996,*Ba1anda224421}

%Ideally, the key ingredients of SG systems are the site disorder, oxygen vacancies, and non-stoichiometry. It is worth mentioning that unconventional SG behavior has also been found in few pyrochlore oxides Y$_{2}$Mo$_{2}$O$_{7}$,\cite{Shinaoka174422} Tb$_{2}$Mo$_{2}$O$_{7}$,\cite{Gaulin3244} and Y$_{2}$Mn$_{2}$O$_{7}$,\cite{Reimers3387} without having any random site disorder.
%In contrast, Y$_{2}$Mn$_{2}$O$_{7}$ is recently theoretically (\textit{ab initio} calculation) claimed to be simple ferromagnet without any SG transition and the spin-orbit coupling associated with Mn$^{4+}$ ion is found to have negligible effect on orbital degeneracy.\cite{Amirabbasi2020}
%For LNMO, the reitveld refinement of the XRD and NPD data ruled out the possibilty of cation mixing and established the high purity of the sample without any chemical disorder. \textbf{Need to be verified by NPD data analysis.}
%Considering the above points, it seems to be highly unlikely for chemically ordered LNMO to have a SG behavior. Although, there could be a substantial amount of competing exchange interactions, which could give rise to significant frustration to facilitate a SG transition.
 
\subsubsection{Memory Effect using Magnetic Relaxation}
\begin{figure}
	\includegraphics [width = \linewidth] {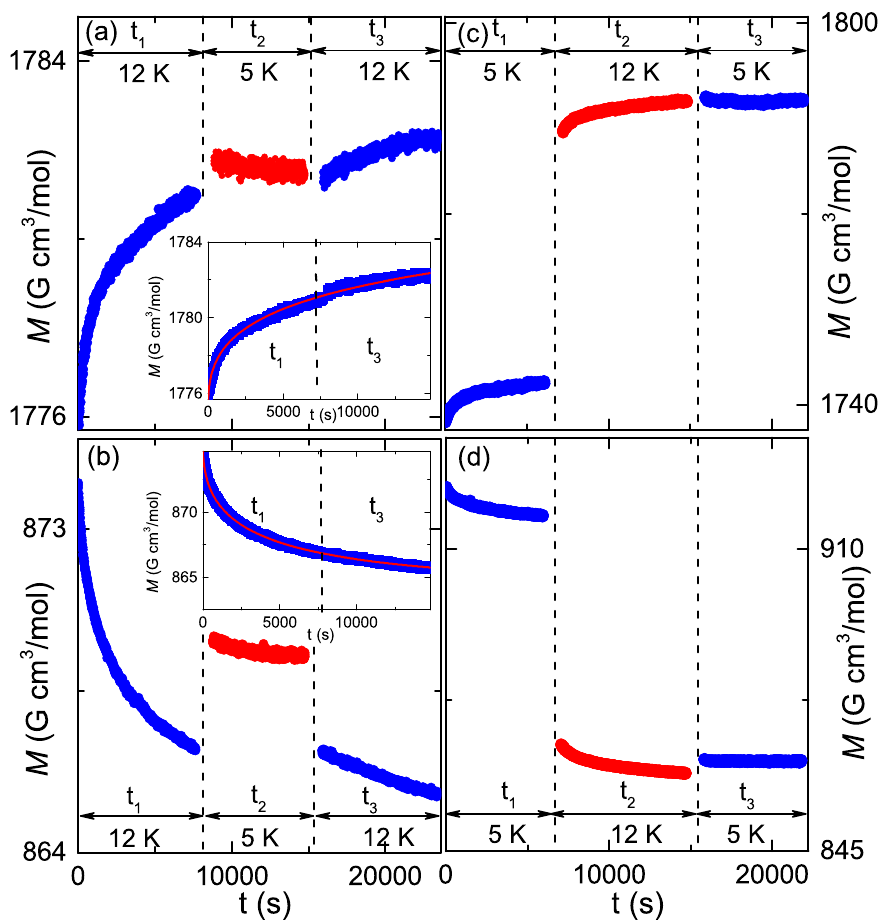}
	\caption{\label{Fig22} Magnetic relaxation measured in negative $T$-cycle for (a) ZFC and (b) FC methods, as described in the text. Insets: Magnetic relaxation at 12~K measured during $t_1$ and $t_3$ for negative $T$-cycle and in ZFC and FC methods. The solid lines are the fits using Eq.~\eqref{strchd eq}. Magnetic relaxation data measured in positive $T$-cycle for ZFC and FC methods are shown in (c) and (d), respectively.}
\end{figure}

To investigate the memory effect in further detail, we have performed the magnetic relaxation measurements following the protocol reported by Sun~$et.~al.$\cite{Sun167206} for both negative and positive-$T$ cycles.

$Negative~T-cycle$: In a negative temperature cycle, we have measured the magnetic relaxation in both ZFC and FC protocols and are plotted in Fig.~\ref{Fig22}(a) and (b), respectively. In the ZFC procedure, the sample was first cooled from 200~K down to 12~K in zero field. At 12~K, a small field of 100~Oe was applied and $M(t)$ was recorded for $t_{1} = 2$~hours, which is found to grow exponentially with $t$. The sample was again cooled down to 5~K in the same magnetic field and $M(t)$ was measured for $t_{2} = 2$~hours. The nature of $M(t)$ curve was found to be almost constant with $t$. Thereafter, the temperature was restored back to 12~K and $M(t)$ was recorded for $t_{3} = 2$~hours in the same field. At this temperature, the $M(t)$ curve was again found to grow exponentially with $t$. As shown in the inset of Fig.~\ref{Fig22}(a), the $M(T)$ data measured during $t_{1}$ and $t_{3}$ follow a continuous growth curve. In the FC process, the sample was cooled in a small magnetic field of 100~Oe down to 12~K. Once it reached 12~K, the magnetic field was switched off and the decay of magnetization with $t$ was measured for $t_{1} = 2$~hours. The sample was further cooled down to 5~K in zero field and $M(t)$ was recorded for $t_{2} = 2$~hours at 5~K. This $M(t)$ curve was found to be almost constant with $t$. Subsequently, the sample was heated back to 12~K in zero field and $M(t)$ was recorded for $t_{3} = 2$~hours at 12~K. The $M(t)$ curve measured during $t_{3}$ was found to decay exponentially with $t$ as a continuation of $M(t)$ curve recorded during $t_{1}$ [see the inset of Fig.~\ref{Fig22}(b)].

Thus, the continuous growth and decay of magnetization during $t_1$ and $t_3$ obtained for the ZFC and FC processes, respectively indicate that the state of the system at 12~K is recovered after a temporary cooling. This a clear demonstration of the memory effect where the system tries to remember the initial state even after going through a change of magnetization. These continuous curves could be fitted well by the stretched exponential function [Eq.~\eqref{strchd eq}] with $\beta \simeq 0.37$ and 0.48 for the ZFC and FC processes, respectively which are consistent with the magnetic relaxation measurements.

$Positive~T-cycle$: Similar to the negative $T$-cycle, we have also done magnetic relaxation measurements in positive $T$-cycle in both ZFC and FC protocols. For ZFC procedure, the sample was first cooled from 200~K down to 5~K in zero magnetic field and $M(t)$ was recorded for $t_{1} = 2$~hours, after applying a small magnetic field of 100~Oe. Then, the sample was heated up to 12~K in the same field and $M(t)$ was measured for $t_{2} = 2$~hours. Thereafter, the sample was again cooled back to 5~K in the same field and $M(t)$ was recorded for $t_{3} = 2$~hours. In the FC procedure, the sample was first cooled down to 5~K in a small magnetic field of 100~Oe. Once the temperature reached 5~K, the magnetic field was switched off and $M(t)$ was recorded for $t_{1} = 2$~hours. By keeping the field zero, the sample was heated to 12~K and $M(t)$ was recorded for $t_{2} = 2$~hours. In zero field, the sample was further brought back to 5~K and $M(t)$ was measured for $t_{3} = 2$~hours. The measured ZFC and FC data are presented in Fig.~\ref{Fig22}(c) and (d), respectively. Unlike the negative $T$-cycle there is no continuity found in the $M(t)$ data measured during $t_{1}$ and $t_{3}$ for both ZFC and FC measurements in the positive $T$-cycle. This clearly suggests that positive $T$-cycling or temporary heating erases the memory and re-initializes the relaxation process in both ZFC and FC methods. Thus, no memory effect is observed when the temperature is restored back to 5~K.

%Generally, the phenomenon of memory effect in SGs systems can be understood in the framework of two theoretical models: the droplet model\cite{Fisher373} and the hierarchical model.\cite{Lefloch647}
The asymmetric response of magnetic relaxation with respect to both negative and positive $T$-cycles is typically observed for both SG and superparamagnetic systems. In SG systems, this behaviour can be explained on the basis of the hierarchical model.\cite{Lefloch647,Sun167206}  Similarly, for superparamagnetic systems, this behaviour is very well explained by Tsoi~$et.~al.$\cite{Tsoi014445} and Bandyopadhyay~$et.~al.$\cite{Bandyopadhyay214410} in the light of simple two-state superparamagnetic model\cite{Sasaki104405}. Thus, the observed memory effects in ZFC and FC protocols for the negative $T$-cycle further justifies the superparamagnetic nature of the system under investigation.

\section{Summary}
We have successfully synthesized the polycrystalline sample of a new modified cubic spinel compound LNMO. It is found to crystallize in a cubic structure with a non-centrosymmetric space group $P4_{3}32$ and exhibits 1:3 cation order of Ni$^{2+}$ and Mn$^{4+}$ ions. Physical property measurements suggest semiconducting nature of the compound which exhibits a long-range ferrimagnetic ordering at $T_{\rm C} \simeq 125$~K. The analysis of the neutron diffraction data reveals a collinear ferrimagnetic spin structure, with magnetic moments aligned along [110] direction. The moments of each Ni$^{2+}$ or Mn$^{4+}$ sublattice are coupled ferromagnetically whereas the inter-sublattice interaction is antiferromagnetic. The compound is still frustrated due to competing AFM and FM interactions which is also the reason for the reduction of frustration ratio ($f$), despite a highly frustrated pyrochlore lattice geometry. The reduced ordered moment of Ni$^{2+}$ or Mn$^{4+}$ ions also indicates the presence of significant frustration.

The critical exponents obtained from the analysis of magnetization data near $T_{\rm C}$ via modified Arrott and Kouvel-Fisher plots fall in the category of 3D XY universality class. The reliability of these critical exponents and the value of $T_{\rm C}$ are further confirmed from the scaling of magnetization isotherms. A reversible MCE with a large value of $\Delta S_m $ and $RCP$ has been observed over a wide temperature range across $T_{\rm C}$ which can be ascribed to the effect of magnetic frustration. Even the critical analysis of field dependent $\lvert\Delta S_{\rm m}^{\rm peak}\lvert$ and $RCP$ curves produce exponents close to the ones obtained from the critical analysis of magnetization. The second order nature of the phase transition has also been confirmed from the universal scaling of $\Delta S_{\rm m}$ and the nature of $n(H,T)$ curve. Our results demonstrate that, LNMO is a promising refrigerant material, where frustration associated with the competing exchange interactions drives MCE.

The magnetic relaxation below $T_{\rm b}$ follows stretched exponential function demonstrating that the system evolves through a number of metastable states. $\chi^{''}(T)$ below $T_{\rm C}$ depicts multiple anomalies, likely due to the effect of magnetic frustration. The hump at $T=T_{\rm f2}$ shows strong frequency dependency with unusual dynamics. It follows a Arrhenius behaviour in the high temperature regime ($T>27$~K), suggesting superparamagnetic behaviour and a power law behaviour in the low temperature regime ($T<27$~K), suggesting SG dynamics at low temperatures. However, the absence of ZFC memory effect below $T_{\rm b} \simeq 120$~K rules out the possibility of SG transition and confirms the superparamagnetic behaviour down to the lowest measured temperature. Indeed, the behaviour of DC $\chi(T)$ measured in FC and ZFC conditions also seems to substantiate the superparamagnetic nature of the compound with blocking temperature $T_{\rm b} \simeq 120$~K. Nevertheless, the critical slowing down behaviour following a power law at low temperature and the absence of ZFC memory contradict each other and warrants further investigations.
%Thus, the critical slowing down behavior following a power law at low temperature demonstrates a SG behavior associated with freezing mechanism but the absence of ZFC memory implies a superparamagnetic behavior associated with blocking mechanism, which contradict each other and warrants further investigations.
%The physics behind spin freezing and spin blocking mechanism is totally different. Spin freezing is a cooperative phenomena arising due to random frustrated competing AFM and FM interactions, whereas spin blocking is a dynamic phenomena arising due to competition between thermal and the magnetic energies of the magnetic entities. Further experiments including third order susceptibility ($\chi_{3}$)\cite{Bajpai637}, muon spin resonance ($\mu$SR) etc on single crystals would be useful to resolve this issue.

\section{acknowledgments}
We would like to acknowledge SERB, India bearing sanction order No. CRG/2019/000960 and BRNS, India bearing sanction Grant No. 37(3)/14/26/2017-BRNS for financial support.

%\bibliography{reff_LNMO}


\end{document}